\documentclass[
  journal=pasa,
  manuscript=research-paper, 
  year=2026,
  volume=XX,
]{cup-journal}

\usepackage{microtype,siunitx,booktabs,amssymb,amsmath,ulem}
\title{Murriyang cryogenic phased array feed: spectral-line results and noise-reduction methods} 

\author{L.~Staveley-Smith}
\affiliation{International Centre for Radio Astronomy Research (ICRAR), University of Western Australia, 35 Stirling Highway, Crawley, WA 6009, Australia}
\email[Lister Staveley-Smith]{Lister.Staveley-Smith@uwa.edu.au}

\author{S.~Barker}
\affiliation{ {ATNF, CSIRO, Space and Astronomy, PO Box 76, Epping, NSW 1710, Australia}}

\author{R.~Berangi}
\affiliation{ {ATNF, CSIRO, Space and Astronomy, PO Box 76, Epping, NSW 1710, Australia}}

\author{A.B.~Bolin}
\affiliation{ {ATNF, CSIRO, Space and Astronomy, PO Box 76, Epping, NSW 1710, Australia}}

\author{S.~Broadhurst}
\affiliation{ {ATNF, CSIRO, Space and Astronomy, PO Box 76, Epping, NSW 1710, Australia}}

\author{J.D.~Bunton}
\affiliation{ {ATNF, CSIRO, Space and Astronomy, PO Box 76, Epping, NSW 1710, Australia}}

\author{N.~Carter}
\affiliation{ {ATNF, CSIRO, Space and Astronomy, PO Box 76, Epping, NSW 1710, Australia}}

\author{S.~Castillo}
\affiliation{ {ATNF, CSIRO, Space and Astronomy, PO Box 76, Epping, NSW 1710, Australia}}

\author{W.~Chandler}
\affiliation{Warren Chandler Pty Ltd., 86 Paterson Rd, Springwood, NSW 2777, Australia}

\author{A.~Chippendale}
\affiliation{ {ATNF, CSIRO, Space and Astronomy, PO Box 76, Epping, NSW 1710, Australia}}

\author{J.R.~Dawson}
\affiliation{School of Mathematical and Physical Sciences and Astrophysics and Space Technologies Research Centre, Macquarie University, 2109, NSW, Australia}
\alsoaffiliation{ {ATNF, CSIRO, Space and Astronomy, PO Box 76, Epping, NSW 1710, Australia}}

\author{F.~Di Dio}
\affiliation{ {ATNF, CSIRO, Space and Astronomy, PO Box 76, Epping, NSW 1710, Australia}}

\author{ {A.R.~Dunning}}
\affiliation{ {ATNF, CSIRO, Space and Astronomy, PO Box 76, Epping, NSW 1710, Australia}}

\author{S.~Gordon}
\affiliation{ {ATNF, CSIRO, Space and Astronomy, PO Box 76, Epping, NSW 1710, Australia}}

\author{J.A.~Green}
\affiliation{ {ATNF, CSIRO, Space and Astronomy, PO Box 76, Epping, NSW 1710, Australia}}
\alsoaffiliation{SKA Observatory, SKA-Low Science Operations Centre, 26 Dick Perry Avenue, Kensington, WA 6151, Australia}

\author{A.~Hafner}
\affiliation{Sydney Institute for Astronomy, School of Physics, University of Sydney, NSW 2006, Australia}

\author{D.B.~Hayman}
\affiliation{ {ATNF, CSIRO, Space and Astronomy, PO Box 76, Epping, NSW 1710, Australia}}

\author{D.~Humphrey}
\affiliation{ {ATNF, CSIRO, Space and Astronomy, PO Box 76, Epping, NSW 1710, Australia}}

\author{A.~Jameson}
\affiliation{Centre for Astrophysics and Supercomputing, Swinburne University of Technology, Mail H39, PO Box 218, Hawthorn, VIC 3122, Australia}
\alsoaffiliation{ARC Centre of Excellence for Gravitational Wave Discovery (OzGrav), Australia}

\author{S.~Johnston}
\affiliation{ {ATNF, CSIRO, Space and Astronomy, PO Box 76, Epping, NSW 1710, Australia}}

\author{J.F.~Kaczmarek}
\affiliation{SKA Observatory, SKA-Low Science Operations Centre, 26 Dick Perry Avenue, Kensington, WA 6151, Australia}
\alsoaffiliation{ {ATNF, CSIRO, Space and Astronomy, PO Box 76, Epping, NSW 1710, Australia}}

\author{J.~Ma}
\affiliation{Xinjiang Astronomical Observatory, Chinese Academy of Sciences, 150 Science 1-Street, Urumqi, Xinjiang 830011, China}

\author{G.~Perry}
\affiliation{ {ATNF, CSIRO, Space and Astronomy, PO Box 76, Epping, NSW 1710, Australia}}

\author{M.~Pilawa}
\affiliation{ {ATNF, CSIRO, Space and Astronomy, PO Box 76, Epping, NSW 1710, Australia}}

\author{J.~Rhee}
\affiliation{International Centre for Radio Astronomy Research (ICRAR), University of Western Australia, 35 Stirling Highway, Crawley, WA 6009, Australia}

\author{L.~Toomey}
\affiliation{ {ATNF, CSIRO, Space and Astronomy, PO Box 76, Epping, NSW 1710, Australia}}

\author{J.~van Aardt}
\affiliation{ {ATNF, CSIRO, Space and Astronomy, PO Box 76, Epping, NSW 1710, Australia}}

\author{N.~Wang}
\affiliation{Xinjiang Astronomical Observatory, Chinese Academy of Sciences, 150 Science 1-Street, Urumqi, Xinjiang 830011, China}

\doi{}

\received {8 Aug 2025}
\revised  {15 Dec 2025}
\accepted {11 Feb 2026}
\published{XX Month 2026}


\begin{document}
\setlength\emergencystretch\columnwidth

\begin{abstract}
Spectral-line results from a new cryogenic phased array feed (cryoPAF) on the Murriyang telescope at Parkes are presented. This array offers a significant improvement in field of view, aperture efficiency, bandwidth,  {chromaticity} and survey speed compared with conventional horn-fed receivers. We demonstrate this with measurements of sky calibrators and observations of 21-cm neutral hydrogen (HI) in the Large Magellanic Cloud (LMC) and the nearby galaxy NGC 6744. Within 0.3 deg of the optical axis, the ratio of system temperature to dish aperture efficiency ($T_{\rm sys}/\eta_{\rm d}$) is 25 K and the ratio with beam efficiency ($T_{\rm sys}/\eta_{\rm mb}$) is 21 K (at 1.4 GHz). For the previously measured $T_{\rm sys} = 17$ K, respective efficiency values $\eta_{\rm d} \approx 0.7$ and $\eta_{\rm mb} \approx 0.8$ are derived.
Our HI observational results are in good agreement with previous results, although detailed comparison with multibeam observations of the LMC suggests that the earlier observations may have missed an extended component of low-column-density gas ($\sim 8\times 10^{18}$ cm$^{-2}$).
We use the cryoPAF zoom-band and wideband data to make a preliminary investigation of whether the large number of simultaneous beams (72) permits the use of novel data reduction methods to reduce the effects of foreground/background continuum contamination and radio-frequency interference (RFI). We also investigate if these methods can better protect against signal loss for the detection of faint, extended cosmological signals such as HI intensity maps.  {Using robust higher-order singular value decomposition (SVD) techniques, we find encouraging results for the detection of both compact and extended sources, including challenging conditions with high RFI occupancy and significant sky continuum structure}. Examples are shown that demonstrate that  {3D SVD techniques offer a significant improvement in noise reduction and signal capture compared with more traditional layered 2D techniques}. 
\end{abstract}

\section{Introduction}
\label{sec:intro}

\begin{figure}[t]
    \centering
    \includegraphics[width=1.0\textwidth]{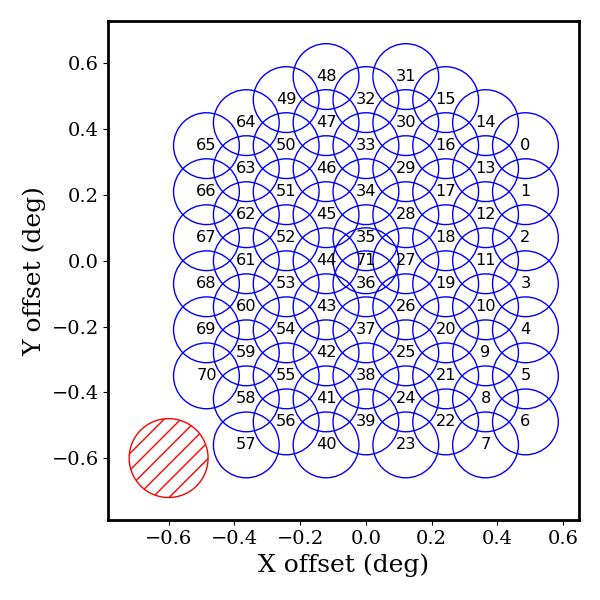}
    \caption{The cryoPAF  {{\tt closepack72} beam footprint defining the arrangement of beams within the field of view}. When the feed angle is set to 0 deg, the X and Y offset correspond to the sky offsets in azimuth and elevation, respectively, with respect to the optical axis of the telescope (beam 71 for this  {footprint}). The radius of each blue circle is 0.1 deg;  {the hatched red
    circle represents the approximate beam size at 1.4 GHz ($\sim 0.24$ deg).}}
    \label{fig:closepack72}
\end{figure}

\begin{figure}
    \centering
    \includegraphics[width=1.0\textwidth]{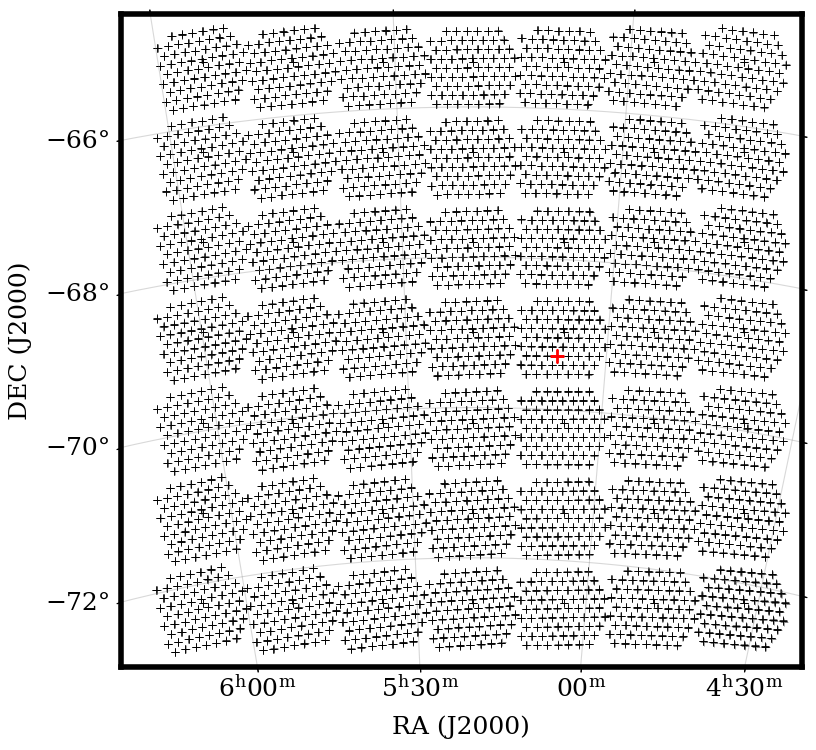}
    \caption{The $7\times7$ pointing pattern used to observe the LMC. There was no parallactification of the cryoPAF during these observations, so the fields are rotated with respect to Figure~\ref{fig:closepack72}. The red cross indicates the position of the archival multibeam spectrum discussed in the text.}
    \label{fig:LMCpointings}
\end{figure}

\begin{figure*}[t]
    \centering
    \includegraphics[width=1.0\textwidth]{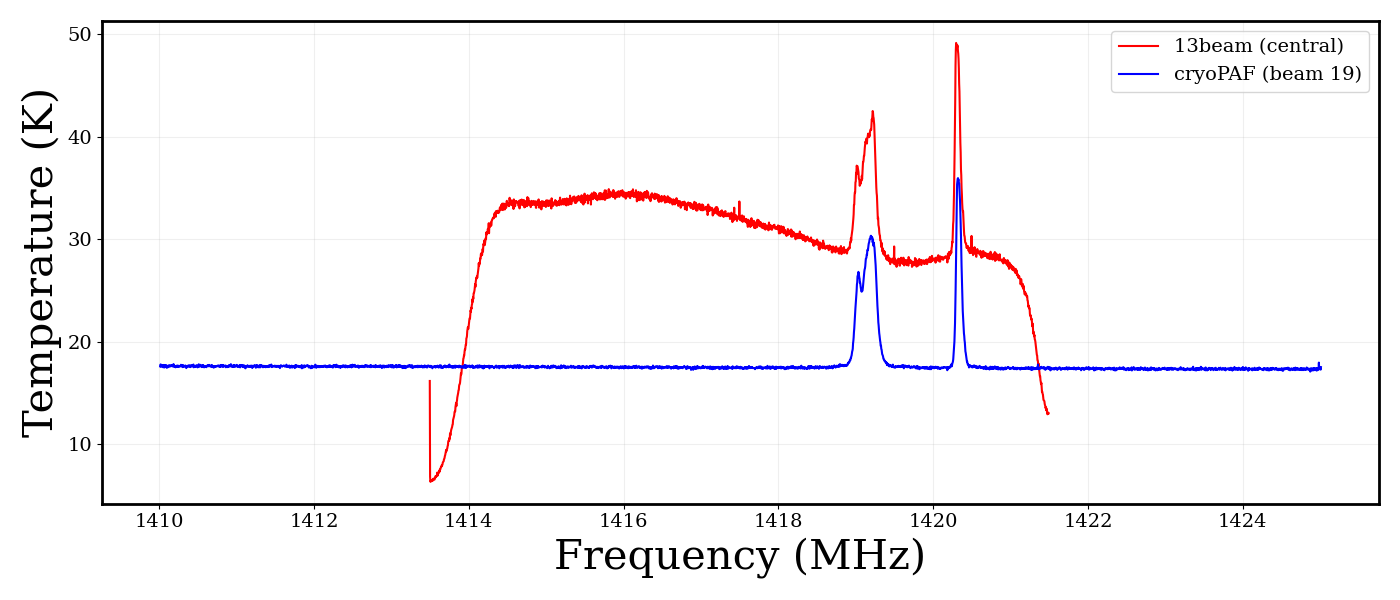}
    \caption{A comparison of HI spectra at a similar position within the LMC (RA = $05^{\rm h}07^{\rm m}10^{\rm s}$, Dec = $-69^{\circ}14'41''$, J2000), as marked with the red cross in Figure~\ref{fig:LMCpointings}. The red spectrum is a 5-s archival P312 multibeam spectrum from 1998 December 16 \citep{2003MNRAS.339...87S}. The blue spectrum is 10 s of data from cryoPAF beam 19 from 2024 November 24, smoothed to a similar resolution as the multibeam data (3.9 kHz). Both are Stokes $I$ spectra. No bandpass calibration or baseline fitting has been applied to either spectrum. The cryoPAF brightness temperature and barycentric frequency scales are approximately correct. The multibeam spectrum has an arbitrary temperature scale, with no barycentric correction applied.}
    \label{fig:comparison}
\end{figure*}

Multifeed radio telescopes have revolutionised radio astronomy over the last few decades. They have expanded the field of view over which sensitive observations can be gathered, with only minor losses in other parameters such as system temperature, off-axis efficiency, bandwidth and polarimetric performance. Examples of such systems at gigahertz frequencies include: the MIT-NRAO 7-beam receiver used to conduct the PMN continuum surveys \citep{1989AJ.....97.1064C}; the Parkes multibeam receiver \citep{1996PASA...13..243S} used to conduct HIPASS \citep{2001MNRAS.322..486B} and the pulsar surveys \citep{2001MNRAS.328...17M} responsible for the discovery of fast radio bursts (FRBs) \citep{2007Sci...318..777L}; the Arecibo L-band Feed Array (ALFA) on the late Arecibo telescope, used to conduct the ALFALFA HI survey \citep{2005AJ....130.2598G}; and the 19-beam array on the Five-hundred meter Aperture Spherical Telescope (FAST) being used to conduct several HI, pulsar and FRB surveys \citep{8331324}. 

More recently, a new type of multifeed technology  {for reflector antennas}, the phased array feed (PAF), has been developed. The main example is the the Australian SKA Pathfinder (ASKAP), which consists of 36 dishes each fitted with a PAF with 36 beams \citep{2021PASA...38....9H}. Notwithstanding the fact that the ASKAP PAFs are held close to ambient temperature, and therefore have a high system temperature of about 75K, ASKAP is conducting world-class continuum, spectral line, polarisation and transient surveys, including RACS \citep{2020PASA...37...48M}, WALLABY \citep{2020Ap&SS.365..118K}, GASKAP \citep{2013PASA...30....3D}, EMU \citep{2025PASA...42...71H} and VAST \citep{2013PASA...30....6M}. It was also responsible for the early localisation of numerous FRBs, leading to the eventual detection of the `missing ordinary matter' in the Universe \citep{2020Natur.581..391M}.

 {Similarly, the APERture Tile In Focus (Apertif) system \citep{2022A&A...658A.146V} was deployed on the Westerbork Synthesis Radio Telescope (WSRT) for a number of years and used to conduct successful northern hemisphere observations of extragalactic HI \citep{2022A&A...667A..38A}, continuum sources \citep{2022A&A...667A..39K} and FRBs \citep{2023A&A...672A.117V}. A prototype cryogenic PAF (AO-19) was developed for the Arecibo telescope \citep{2014SPIE.9147E..9QC}, with plans for a more capable 69-element Advanced Cryogenic L-band Phased Array Camera for Arecibo (ALPACA) in progression before the demise of the telescope \citep{9232378}. However, the first cryogenic phased array feed used in astronomy was the 19-element, 7-beam Focal L-band Array for the GBT \citep[FLAG; ][]{2018AJ....155..202R}, which was commissioned for HI observations by \citet{2021AJ....161..163P}.}

The advantages of the using PAFs, rather than traditional feed horns, in multifeed systems are numerous: (1) they allow for correction of off-axis distortion and gain loss  {\citep{2012ITAP...60..903E}}; (2) they illuminate the dish with  {high efficiency \citep{2004ExA....17..149I}}; (3) they have the ability to Nyquist sample the focal plane (in general, traditional feed horns cannot be spaced by less than 2 beamwidths before gain loss sets in)  {\citep{2000SPIE.4015..308F}}; (4) beam shape and sidelobes can be adjusted, for example to reduce sidelobes, or avoid radio frequency interference (RFI)  {\citep{2008ISTSP...2..635J}}; (5) the fractional frequency range is higher than for conventional feed horns  {\citep{2004ExA....17..149I}}; and (6) they have lower  {`standing waves'} as a result of reduced reflection from the focal plane \citep{2017PASA...34...51R}. Similar advantages apply to filled aperture arrays, especially at sub-gigahertz frequencies, which are basically PAFs placed on the ground and facing upwards. The main example is the high-band antenna system of LOFAR \citep{2013A&A...556A...2V}.

In an attempt to combine the advantages of PAFs with the low system temperatures of conventional receivers, CSIRO (co-funded by the ARC and a consortium of universities and institutes) has built a next-generation cryogenic PAF (hereafter cryoPAF) for the Murriyang radio telescope at Parkes. The main science drivers for cryoPAF construction were FRBs and HI intensity mapping, but the instrument was designed and constructed as an open facility instrument to replace the original multibeam receiver, and extend its capabilities, particularly in frequency range and field of view. In summary, the cryoPAF operates in the frequency range 0.7 to 2 GHz with up to 72 simultaneous dual-polarisation beams  {at a spectral resolution as high as 274 Hz}. The on-dish system temperature is about 17 K at 1.4 GHz, similar to state-of-the-art single-pixel radio telescopes such as MeerKAT \citep{2016mks..confE...1J}.

A brief technical description of the cryoPAF is given in \citet{10238280}, and the full system design and description (front end and signal processing) is in preparation. The purpose of this paper is to report on commissioning of the spectral-line aspects of the receiver, concentrating on observations of local galaxies and a description of new aspects of spectral-line data reduction methodology, with particular reference to the future detection of weak cosmological signals.

The extraction of weak cosmological signals received by radio telescopes can be severely compromised by myriad forms of contamination, including excess noise from receivers and the ground, distortions due to the troposphere and ionosphere, unwanted radiation from Galactic and extragalactic sources, and radio frequency interference (RFI). The latter contaminant is especially severe at gigahertz frequencies and below, due to the proliferation of communication devices. The radio-quiet zones around the new generation SKA telescopes and pathfinders \citep[e.g.][]{6901992} have greatly helped reduce the worst of RFI problems, but many existing telescope locations are less protected. Furthermore, the proliferation of space-based communications platforms and the Global Navigation Satellites System (GNSS) mean that there is no truly radio quiet location on the planet. Also severe at gigahertz frequencies, and below, is non-thermal radiation from Galactic and extragalactic sources which can easily dominate receiver noise. 

Avoidance of frequency ranges 
affected by the worst RFI, or parts of the sky with strongest radio emission is the primary form of mitigation. After that, active outlier removal \citep[e.g][]{refId0} to suppress RFI contamination, and careful calibration \citep{2021ExA....51..193T} to better enable continuum subtraction, are normally necessary. However, despite these measures, cosmological signals such as the Cosmic Dawn and the Epoch of Reionisation (EoR) are extremely weak and manifest themselves in well-defined frequency ranges where contamination from residual RFI and foreground/background continuum sources may still be orders of magnitude stronger. 

In spectral-line radio astronomy, statistical techniques can be employed to remove noise, or extract signal from  {noisy data}. For example, averaging data in time may allow weak RFI signals to be detected and excised, and averaging in frequency may allow residual continuum artifacts to be revealed and excised. Signals may also be better detected using matched filters, for example by adding information about their likely characteristics  {(e.g.\ in frequency, frequency width, polarisation or spatial extent). In the case of transient or periodic sources, signals may also be better detected by averaging or filtering in other dimensions, such as dispersion measure or period.}

In this paper, we use the new cryoPAF commissioning data to examine the usefulness of unsupervised statistical techniques for noise reduction and weak signal extraction.  {Single-dish telescopes normally collect a single spectrum at each observing position. However, the cryoPAF simultaneously collects spectra at up to 72 adjacent positions, with no gaps. Combined with the usual `on-the-fly' scanning technique to map large areas of sky, this means that datasets 
can have high dimensionality}.
However, methods to de-noise or flatten the foreground/background in galaxy or cosmological mapping experiments usually operate in low-dimensional space. Examples include: polynomial fitting in the frequency domain \citep{Wang_2006}; Fourier-domain window functions \citep{2012ApJ...745..176V}; principal component analysis \citep[PCA;][]{Zuo_2023}; singular value decomposition \citep[SVD;][]{2021RAA....21...30L}; Kernel PCA \citep{2021MNRAS.508.3551I}; generalised morphological component analysis \citep[GMCA;][]{2020MNRAS.499..304C}; independent component analysis \citep[ICA;][]{761722,2012MNRAS.423.2518C}; Gaussian processes \citep{2018MNRAS.478.3640M}; scattering transforms \citep{2022A&A...668A.122D}; wavelets \citep{2024A&A...681A...1A}; and, more recently, deep learning \citep{2021JCAP...04..081M, 2024MNRAS.527.3517M, 2024MNRAS.532.2615C}. In the context of SKA cosmology, comparisons between some of these methods are given in \cite{2015MNRAS.447..400A}, \cite{2015aska.confE...5C}, \cite{2022MNRAS.509.2048S} and \cite{2025MNRAS.543.1092B}.

Most of these methods have been implemented in flattened 1D or 2D space (e.g.\ frequency, frequency--frequency, or frequency-sky pixel space). Deep learning techniques are well suited to data of higher dimensionality, but require extensive training using advance knowledge of the characteristics of cosmological signals, foregrounds/backgrounds, and RFI.
As a step towards de-noising data of high dimensionality, in this paper we explore higher-order/tensor versions of SVD to quantify the improvement in compact and extended source removal over 1D or 2D approaches, using linear algebra.  {Tensor SVD  is itself closely related to deep learning techniques, but has the advantages of being unsupervised and fast. Disadvantages are similar to those of standard SVD \citep{golub2013matrix} such as only being able to deal with linear systems, and sensitivity to scaling of data.}

In this paper, we  {summarise cryoPAF beamformer weight calibration for early commissioning work in Section~\ref{sec:beam_weights} and} describe the cryoPAF spectral-line commissioning observations, data analysis and basic science verification in Section~\ref{sec:observations}. We then discuss possible future techniques, based on SVD and higher-order SVD in Section~\ref{sec:techniques}. Results from the application of these techniques to cryoPAF commissioning data are described in 
Section~\ref{sec:results}. There is further discussion in Section~\ref{sec:discussion}, and we conclude in Section~\ref{sec:conclusions}. 
For calculating cosmological distances, we assume Planck 2018 cosmological parameters $(h, \Omega_{\rm m}) = (0.674, 0.315)$ \citep{2020A&A...641A...6P}, and $\Omega_{\Lambda} = 1-\Omega_{\rm m}$. For comparison with previous work, wavenumbers are quoted in units of $h/$Mpc.

\section{Observations and analysis}
\label{sec:observations}

\subsection{Beamformer weights}
\label{sec:beam_weights}

 {Maximum signal-to-noise ratio (maxSNR) weights for the cryoPAF's digital beamformer were calculated following a procedure similar to that described in \cite{2021PASA...38....9H} for ASKAP's PAFs.  The cryoPAF beamformer calculates and accumulates the array covariance matrix (ACM) $\mathbf{R}$ that is formed from the pairwise correlation of all PAF ports.  The ACM was measured both towards $\mathbf{R}_\text{on}$ and away from $\mathbf{R}_\text{off}$ a strong point source.  The response of the array in the direction of the point source is then obtained from the dominant eigenvector $\mathbf{u_1}$ of $\mathbf{R}_\text{on} - \mathbf{R}_\text{off}$ \citep{2008ISTSP...2..635J}.  The maxSNR beamformer weights \citep{LoLee_OptimizationDirectivity, Applebaum_AdaptiveArrays} were then calculated by 
\begin{equation}
    \mathbf{w} = \mathbf{R}_\text{off}^{-1}\mathbf{u}_1.
\end{equation}}
 
  {The current work uses ACMs measured on Virgo~A on 2024 November 7 to calculate beamformer weights. $\mathbf{R}_\text{off}$ was measured once on an assumed empty patch of sky, offset from Virgo~A by \ang{-1.5} in cross-elevation. $\mathbf{R}_\text{on}$ was then measured at each of 72 pointing offsets about Virgo A, positioning the source at each offset location in the sky towards which a beam is desired to be directed to form the {\tt closepack72} beam footprint of Figure \ref{fig:closepack72}. Three 10-s 
  ACM accumulations were recorded for each pointing with a frequency resolution of \qty{133.3}{\kilo\hertz}. Beamformer weights were calculated at the same frequency resolution as the ACMs after which the algorithm described by \cite{Chippendale_InterferenceMitigation} was applied to interpolate weights at frequencies where the signal from Virgo~A had been drowned out by RFI.  The weights were then linearly interpolated to the frequency resolution of \qty{14.8}{\kilo\hertz} at which digital beamforming is performed.
}

 {For the current work, beamformer weight calculation was performed independently for X and Y polarisations. Weights for X-polarisation beams only included X-polarisation PAF ports, and weights for Y-polarisation beams only included Y-polarisation PAF ports.  This was enforced by extracting sub-matrices from each ACM containing only correlations between ports of matching polarisation and repeating the weight calculation process independently for both XX and YY submatrices.  This is convenient for early commissioning as the polarisation of the beam is well defined by the polarisation of the array elements.  In the future, improved sensitivity and polarisation performance may be obtained by applying polarisation calibration \citep{WijnholdsPolarimetry, WarnickPolarimetry} to weights calculated from the full ACM, including all co-polar and cross-polar correlations.}

\begin{figure}[t]
    \centering
    \includegraphics[width=1.0\textwidth]{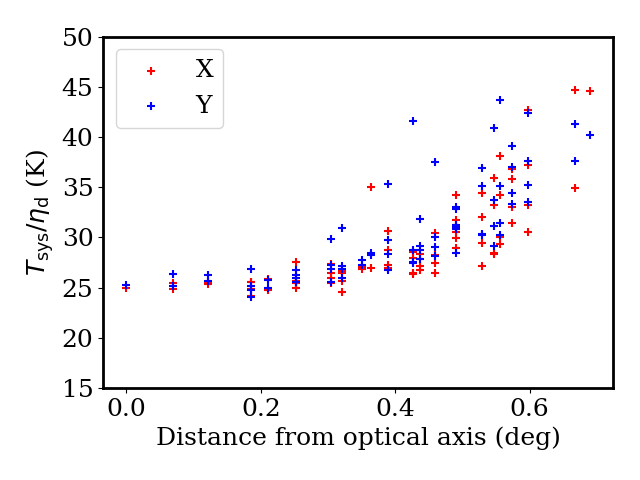}
    \includegraphics[width=1.0\textwidth]{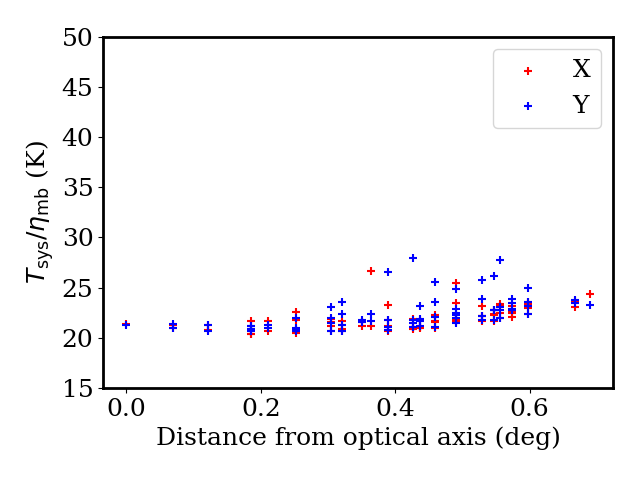}
    \caption{System temperature measurements  {taken on 2024 November 18} for all 72 cryoPAF beams and both orthogonal polarisations at 1.4 GHz from calibrations using (top) the flux density calibrator PKS B1934-638 and (bottom) the Galactic HI source S9. The B1934-638 measurement includes a dish efficiency term ($\eta_{\rm d}$) which decreases away from the optical axis. The S9 measurement includes a main beam efficiency term ($\eta_{\rm mb}$), which decreases less quickly away from the optical axis.}
    \label{fig:calibration}
\end{figure}

\begin{figure}
    \centering
    \includegraphics[width=1.0\textwidth]{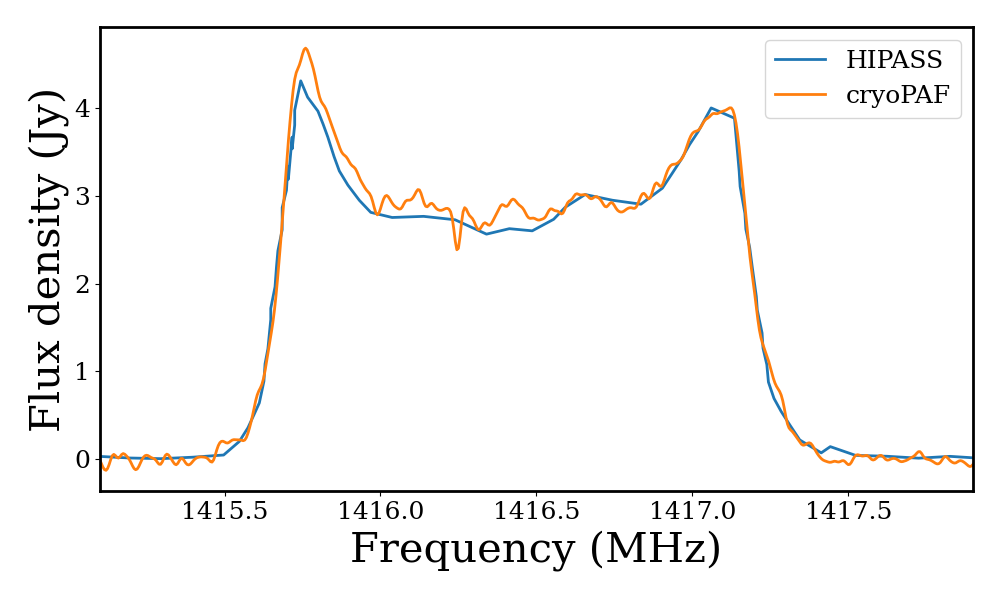}
    \caption{A spatially integrated cryoPAF HI spectrum of NGC 6744, compared with an integrated spectrum from the HIPASS Bright Galaxy Catalogue \citep{2004AJ....128...16K}. 
    The cryoPAF spectrum has been Hanning smoothed to a resolution of 75 kHz; the HIPASS resolution is 85 kHz.}
    \label{fig:NGC6744-comparison}
\end{figure}

\begin{figure*}
    \centering
    \includegraphics[width=0.47\textwidth]{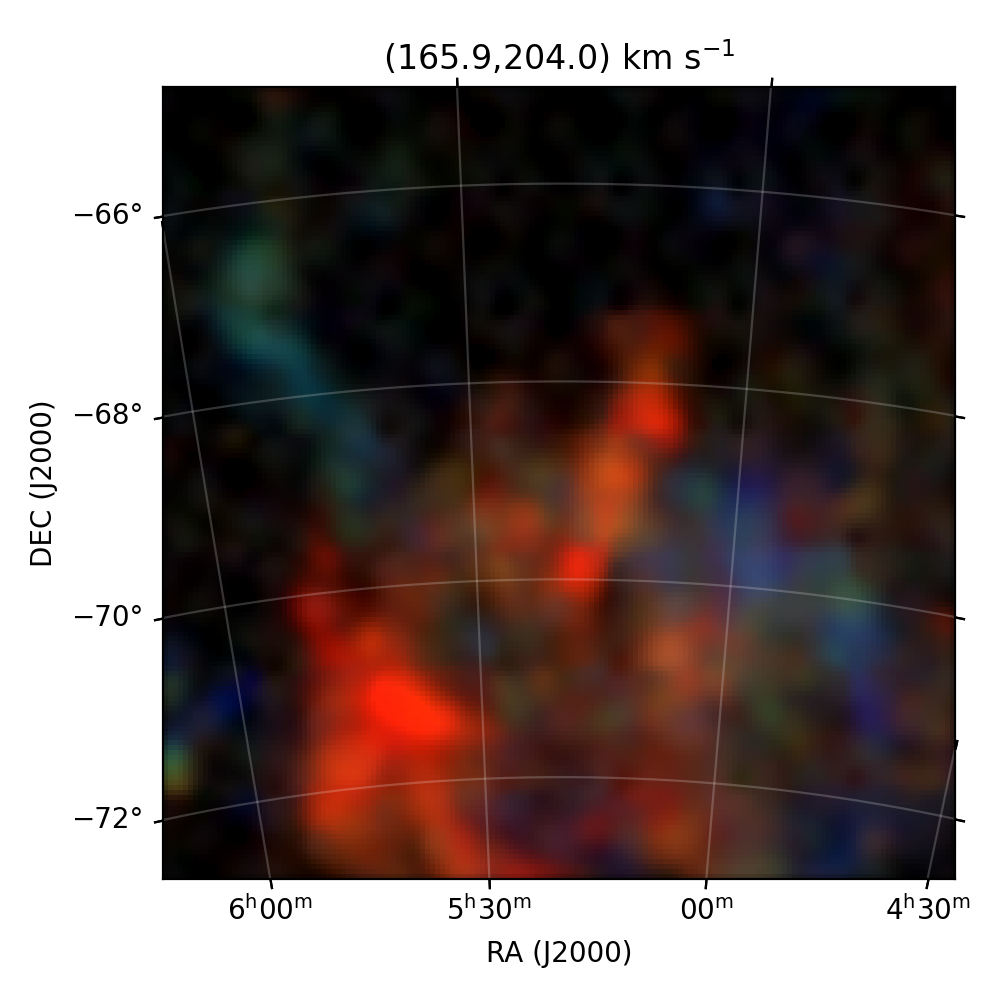}
    \includegraphics[width=0.47\textwidth]{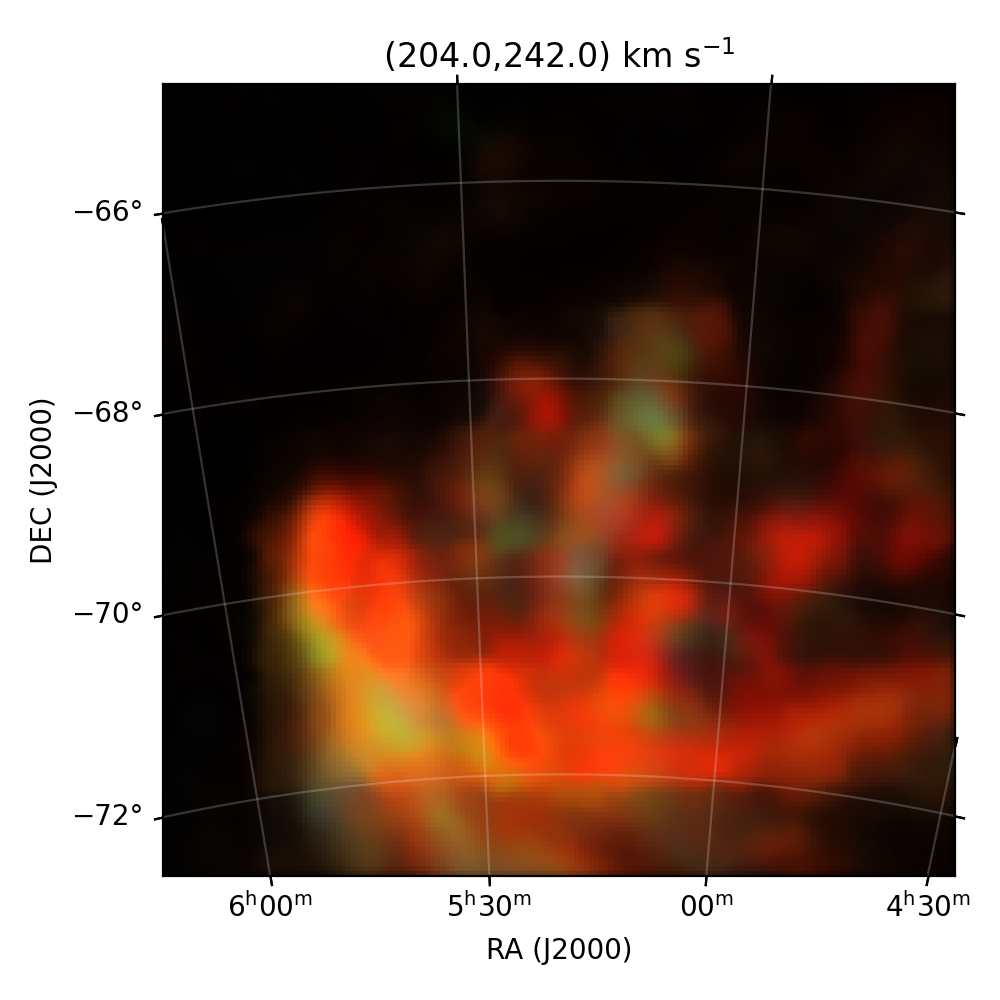}\\[-1cm]
    \includegraphics[width=0.47\textwidth]{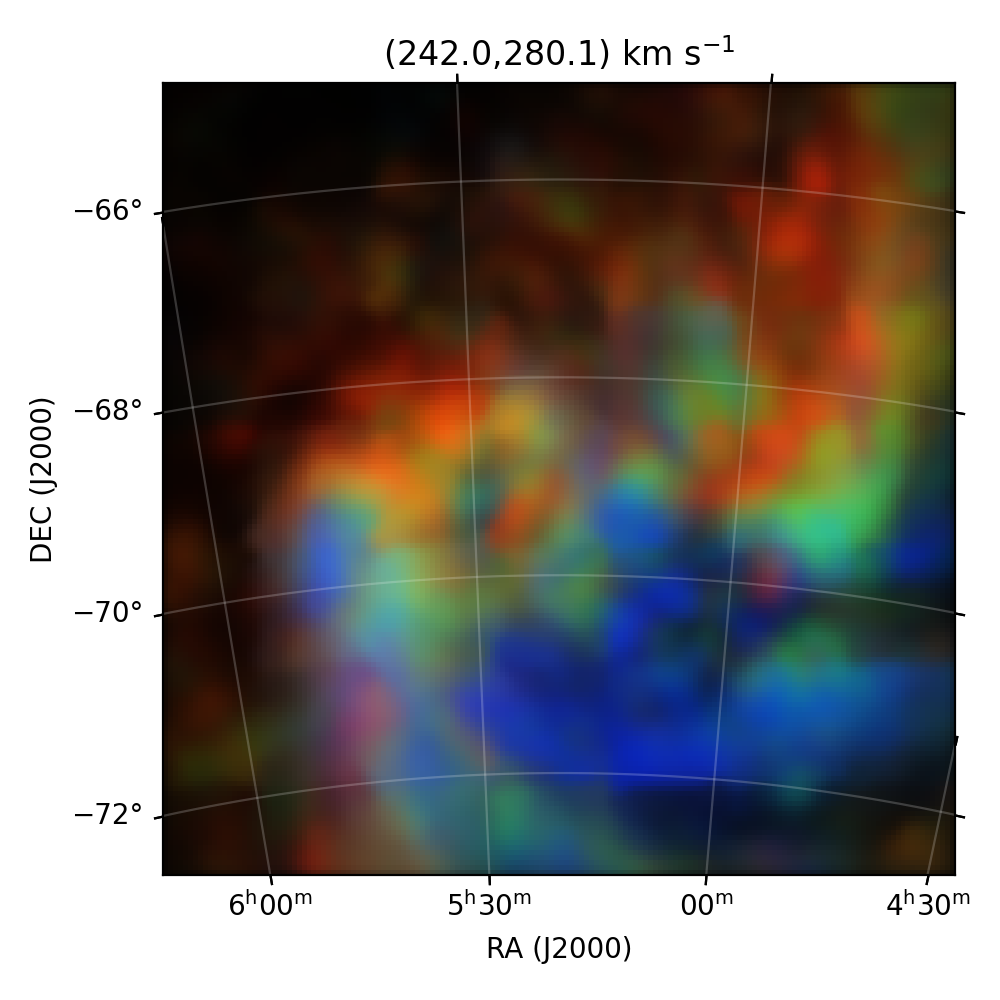}
    \includegraphics[width=0.47\textwidth]{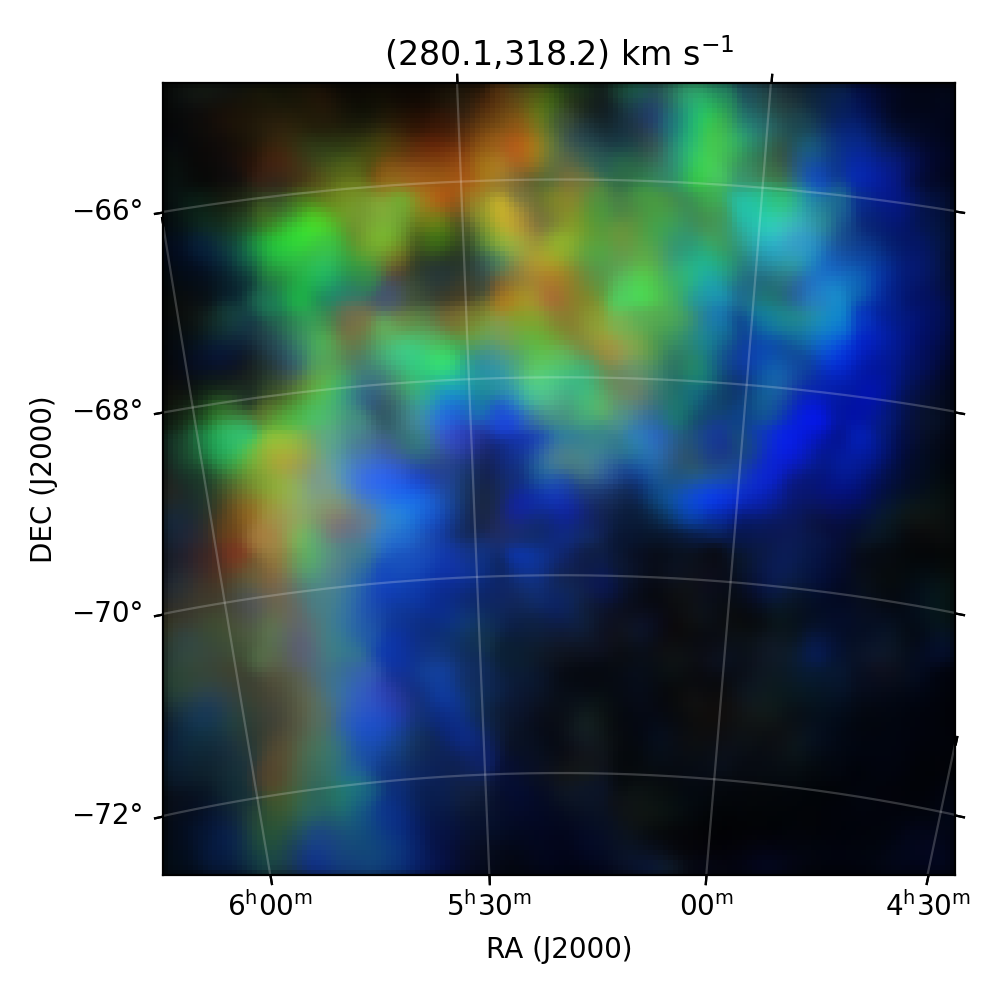}\\[-1cm]
    \includegraphics[width=0.47\textwidth]{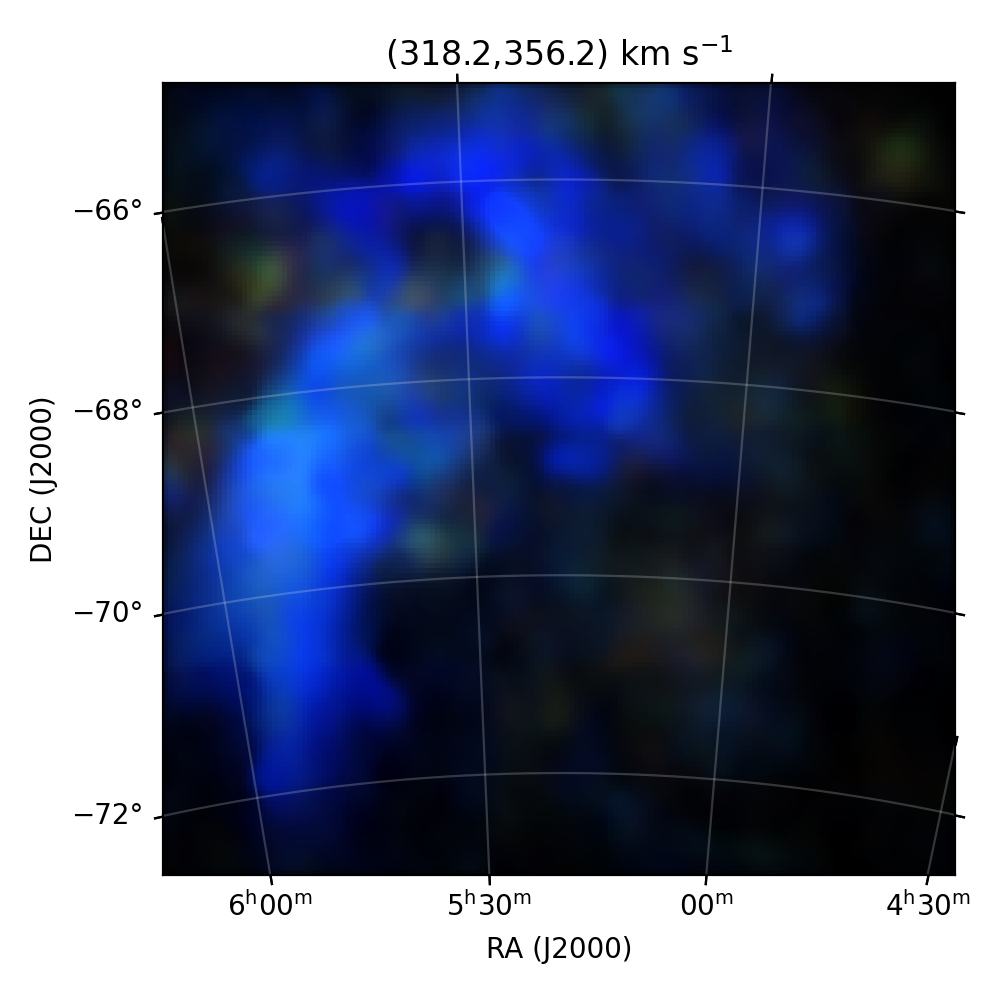}
    \includegraphics[width=0.47\textwidth]{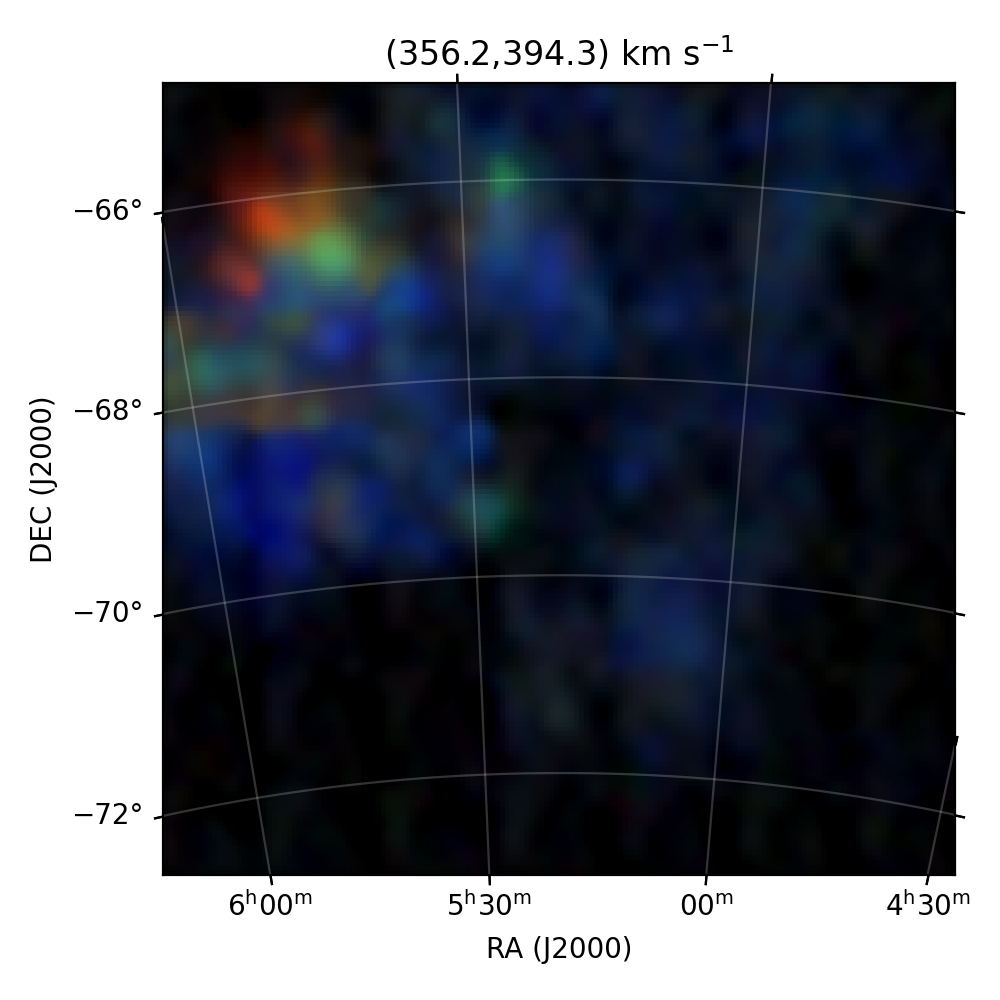}    
    \caption{RGB images of the LMC from 165.9 to 394.3 km~s$^{-1}$ (barycentric) in chunks of width 38 km~s$^{-1}$. Each colour channel has a width of 12.7 km~s$^{-1}$, with the blue channel starting at the lowest velocity in each image. The individual images are scaled to peak brightness temperatures of 0.4, 4.8, 19.5, 27.8, 13.5 and 0.4 K, respectively, with a stretch of 0.5.}
    \label{fig:LMC_channel_maps}
\end{figure*}

\begin{figure*}[t]
    \centering
    \includegraphics[width=0.6\textwidth]{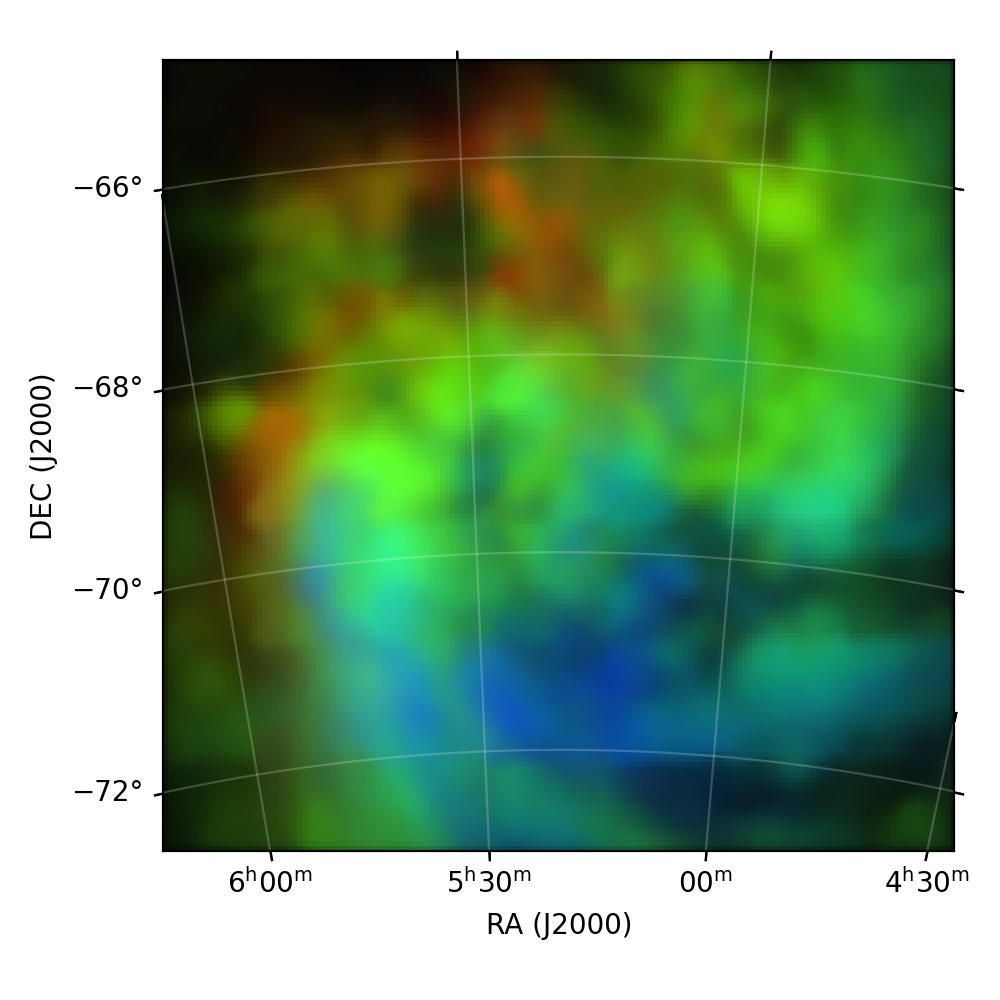}\\[-5mm]
    \caption{A cryoPAF RGB column density image of HI in the LMC, coloured by velocity over the barycentric frequency range 1418.5 to 1419.5 MHz (191 to 403 km~s$^{-1}$). The maximum column density is $5\times 10^{21}$ cm$^{-2}$.}
    \label{fig:LMC_column_density}
\end{figure*}

\begin{figure}
    \centering
    \includegraphics[width=1.0\textwidth]{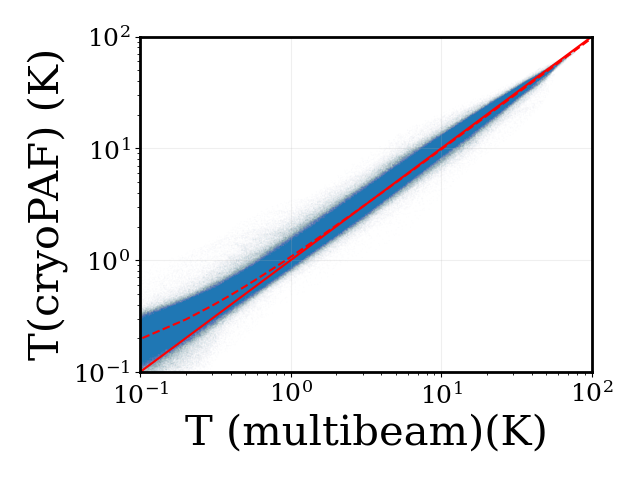}
    \caption{A pixel-pixel comparison of HI brightness temperature in the LMC as measured from RA-Dec-velocity data cubes from the multibeam \citep{2003MNRAS.339...87S} and the cryoPAF. The spatial coverage for the comparison is the same as shown in Figure~\ref{fig:LMC_column_density}; the frequency coverage was 1 MHz (211 km~s$^{-1}$). The two data cubes were position- and resolution-matched. At a frequency resolution 5 kHz (1.1 km~s$^{-1}$), the corresponding rms of the datasets is approximately 22mK and 18mK, respectively. The solid red line represents equality of temperature scales; the dashed red line accounts for the excess emission ($\sim 100$mK) seen in the cryoPAF data, as well as a small offset ($\times 0.97$) in the temperature scale.}
    \label{fig:lmc-comparison}
\end{figure}

\begin{figure*}[t]
    \centering
    \includegraphics[width=0.49\textwidth]{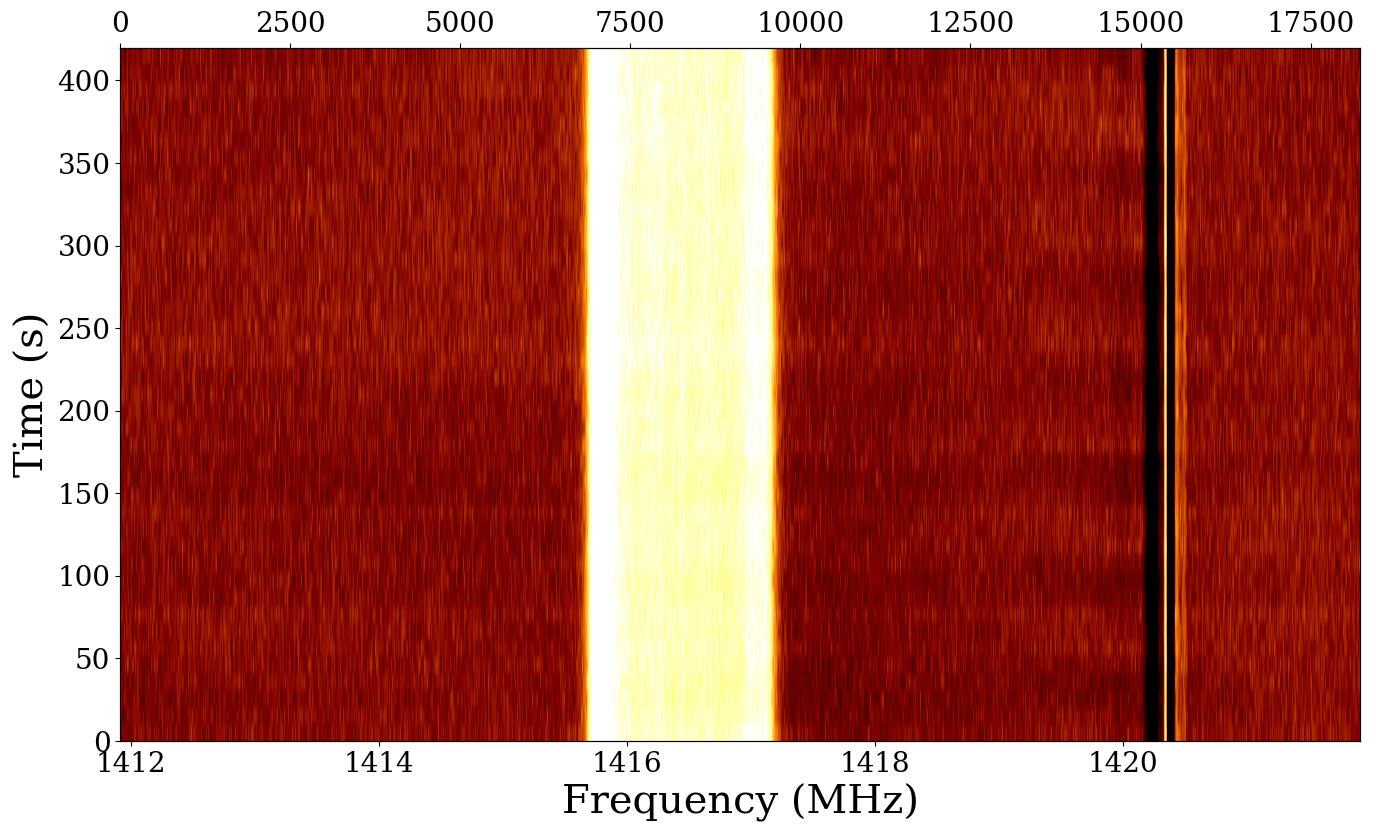}
    \includegraphics[width=0.49\textwidth]{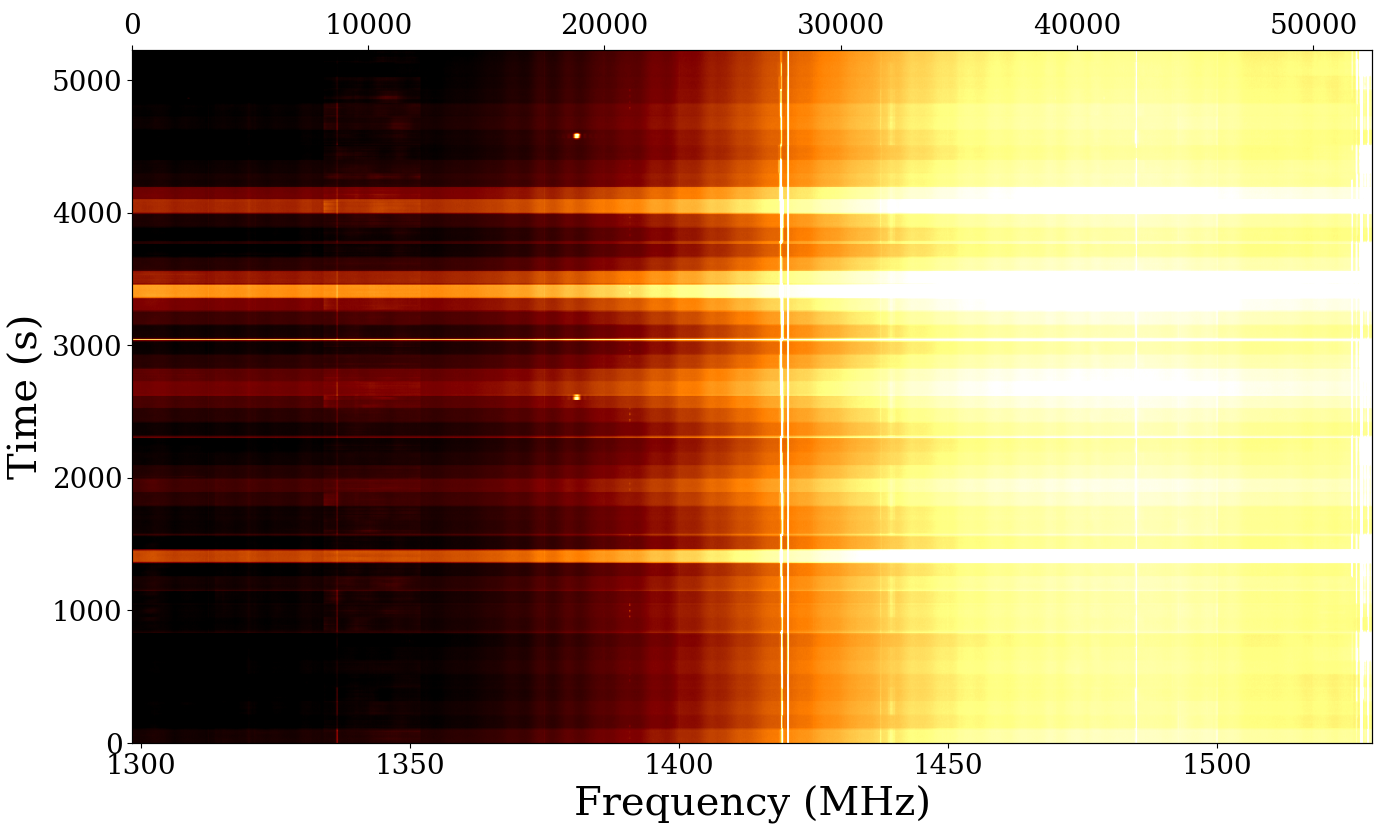}

    \caption{Frequency-time waterfall plots of Stokes $I$ spectral data taken with the central beam  {(beam 71)} of the Murriyang cryoPAF. {\it Left}: single-pointing zoom band data taken on 2025 February 24 whilst observing the NGC 6744. An off-source reference spectrum was applied to remove bandpass ripple. The wide vertical stripe is redshifted HI emission from NGC 6744; the narrow positive and negative lines at 1420.4 MHz are due to Galactic HI at the on-source and off-source positions. The only obvious artefacts in the data are the faint horizontal stripes spaced every minute, which are due to short-term gain variations. 
    {\it Right}: wideband data taken on 2024 November 18 whilst scanning the LMC. This unedited and uncalibrated data shows vertical artefacts due to RFI and  {bandpass} ripple,  and horizontal lines due to strong continuum sources in the LMC. Brief bursts from the GPS L3 beacon at 1380 MHz are apparent. The 21-cm line emission from the LMC and Galaxy are apparent near 1420 MHz. The overall intensity gradient as a function of frequency represents the uncalibrated bandpass shape.
    For both panels, the upper horizontal axis is spectral channel number.}
    \label{fig:waterfall1+2}
\end{figure*}

\subsection{Target observations}

CryoPAF commissioning data for this paper  {were} taken using the Murriyang telescope on 2025 February 24 for the nearby galaxy NGC 6744, and on 2024 November 18 for the Large Magellanic Cloud (LMC). The NGC 6744 data  {consist} of a 7-min azimuth-elevation track, centred on the galaxy, with all 72 beams. The LMC data  {consist} of 66 min of azimuth-elevation tracks of 49 discrete pointing centres with all 72 beams  {(i.e.\ $\sim 80$ sec per pointing). The time resolution for all spectra was 10 sec.} 
 {Although not usually required for cryoPAF data, `off-source' tracks were also recorded.  The reason for this is that the digital `flattening' filter was not fully functional, particularly for the NGC 6744 observations, so a bandpass de-ripple correction was required}. 
Continuous noise source calibration was not enabled at the time of the observations, so contemporaneous scans of S9 \citep{1973A&AS....8..505W,2005A&A...432...45B} and PKS B1934-638 were also made for brightness temperature and flux density calibration purposes, respectively.

All observations were taken in `camera' mode where the telescope observed each pointing centre using the {\tt closepack72} beam  {footprint} (see Figure~\ref{fig:closepack72}), with a pitch of 0.14 deg.  {For a beamwidth of $\sim 0.24$ deg at 1.4 GHz, this pitch satisfies the hexagonal Nyquist criterion \citep[e.g.][]{1989ApJ...343...94S}}. Data  were continuously recorded, so data taken during the brief drives between  {LMC} pointing centres was valid, but flagged as `off source'. 
The overall sky sampling for the LMC observations is shown in Figure~\ref{fig:LMCpointings}. Due to the parallactic angle of the observations, the beam  {footprint} is rotated with respect to Figure~\ref{fig:closepack72} and the gaps between the pointings was greater than intended. 
These gaps (up to $\sim 0.3$ deg in the 36 interior spaces) were interpolated across. For the B1934-638 and S9 calibrations, the telescope was shifted 72 times to place each beam on the source in turn.

The selected bandwidth was 230 MHz over the approximate frequency range 1300 to 1530 MHz, with a selected frequency resolution of 4.4 kHz. Within that band, there was a zoom band covering $1417\pm 5$ MHz for NGC 6744 and $1417.5\pm7.5$ MHz for the LMC. The zoom band frequency resolution was selected to be 0.55 kHz (0.12 km s$^{-1}$). 

Observational data and metadata for the different sources and calibrators were recorded in separate beam files in {SDHDF} format \citep{2024A&C....4700804T}. 
 {In general, hierarchical data format\footnote{https://www.hdfgroup.org} (HDF) \citep{2011hoos.book.....K} has the advantage over more traditional astronomical formats such as FITS of being fast, widely supported, stable, cross-platform, self-documenting, compressible, writeable, modifiable, scalable (supports very large files) and supports inhomogeneous data, including metadata.}
For convenience in later data reduction, the beam files were concatenated into single 72-beam files, also in {SDHDF} format.  {The resultant file for each observation therefore contains data and metadata for each beam and each spectrum (i.e.\ the full-bandwidth, low-resolution spectrum and the zoom band spectrum)}. During commissioning, the equatorial coordinates were only recorded for the central beam (beam 71) at the start of the observation.  {Therefore positions for all beams were recalculated using the UT, the telescope azimuth/elevation, and the beam footprint offsets of Figure \ref{fig:closepack72} rotated for the parallactic angle of the given observation direction}. Due to pointing offsets, this may result in small positional errors.

An example HI spectrum within the LMC zoom band is shown in Figure~\ref{fig:comparison}, together with a spectrum taken at a similar position (shown by the red cross in Figure~\ref{fig:LMCpointings}) using the former multibeam array \citep{2003MNRAS.339...87S}, extracted from the Australia Telescope Online Archive.\footnote{https://atoa.atnf.csiro.au} Neither spectrum has been bandpass calibrated, but the cryoPAF spectrum is clearly much flatter owing to the absence of any narrow-band  {analogue} bandpass filter.

\subsection{Calibration}
\label{sec:calibration}

For each beam, the on-source flux density (or brightness temperature in the case of S9), was measured when the calibration source was within 3 arcmin, and the telescope was `in-lock'  {(i.e.\ correctly slaved to the Murriyang master equatorial telescope)}. 
 {For the unresolved flux calibrator} B1934-638, the `off-source' region was defined using data between 0.5 and 2 deg off source. For the S9 Galactic HI calibration source, `off source' was defined by a straight-line fit to the spectral baseline in a small frequency range outside of the HI emission velocities. Observations for both calibrators were taken on 2025 February 23/24 and 2024 November 18.

The calibration also allows us to make two different measurements of system temperature ($T_{\rm sys}$): (1) its ratio with dish (antenna) aperture efficiency ($\eta_{\rm d}$); and (2) its ratio with main-beam efficiency ($\eta_{\rm mb}$). The former ($T_{\rm sys}/\eta_{\rm d}$) is related to the system-equivalent flux density ($S_{\rm sys}$), which is the source strength in Janskys required to double the system temperature:

\begin{equation}
\frac{T_{\rm sys}}{\eta_{\rm d}} = 10^{-26} \frac{S_{\rm sys} A}{2k_{\rm B}} ,
\label{eq:sefd}
\end{equation}
where $A$ is the geometric collecting area of the antenna and $k_{\rm B}$ is the Boltzmann constant.  {The main-beam efficiency is the ratio of the power received in the main beam (or lobe) of the antenna to the total power received from an extended source of uniform brightness temperature} \citep{2013tra..book.....W}. 

The results are shown in Figure~\ref{fig:calibration} for the 2024 November measurements. Within 0.3 deg of the optical axis, the average values are $T_{\rm sys}/\eta_{\rm d} = 25.5$ K and $T_{\rm sys}/\eta_{\rm mb} = 21.1$ K. The former is close to the  {corresponding} value (24 K) measured by \cite{10238280} at the same frequency. \cite{10238280} also measured $T_{\rm sys}=17$ K, implying a dish efficiency $\eta_{\rm d} \approx$ 70\% near the optical axis and a main-beam efficiency \citep[defined by the extent of S9, which is approximately a degree;][]{1973A&AS....8..505W} $\eta_{\rm mb} \approx$ 80\%. 
 {\citet{2018AJ....155..202R} report a similar values for $T_{\rm sys}/\eta_{\rm d}$ with the GBT FLAG instrument.} 

 {As noted in Section~\ref{sec:beam_weights}, beamformer calibration was only performed in the 2024 November session. The measured system temperatures and beam shapes were therefore poorer in 2025 February due to changes in the complex gain of the analogue electronics.}

The frequency axes for all beams were transformed into the solar system barycentric frame, without regridding of the data. This was conveniently made possible within the {SDHDF} format, which stores a complete vector of the topocentric frequencies for each sub-band for each observation. 

Flux density and brightness temperature calibration was applied to the NGC 6744 and LMC data, respectively. Off-source spectra were used for bandpass calibration of the NGC 6744 data. For the LMC data (see Figure~\ref{fig:comparison}), only a straight-line spectral fit was required to remove the receiver bandpass and continuum sources.

\subsection{Science verification}
\label{sec:verification}

Although only a single pointing was obtained for NGC 6744, it covered the extent of the galaxy, so the data were gridded into a small data cube. The pixel size was 0.6 arcmin in RA and Dec, and 5 kHz (1.1 km~s$^{-1}$) in frequency (velocity). The gridding kernel was a circular Gaussian of  {FWHM} 11.3~arcmin, truncated at a radius of 12~arcmin. The resultant angular resolution (taking into account the Murriyang beam)  {was estimated to be about 18 arcmin}. Five edge-beams with either a high system temperature or a reference  {(off-source) beam which contained HI emission} were discarded prior to gridding. The resultant spectrum, integrated over the whole map, is shown in Figure~\ref{fig:NGC6744-comparison}. Excellent agreement can be seen with the integrated spectrum from the HIPASS Bright Galaxy Catalogue \citep{2004AJ....128...16K}.
The catalogued HIPASS flux integral is $1031\pm57$ Jy~km~s$^{-1}$ or $4.88\pm 0.27$ Jy~MHz; the cryoPAF integral is $5.07\pm 0.25$ Jy~MHz. The values are well within the errors,  {possibly} because the flux uncertainty for resolved sources is normally dominated by knowledge of the beam shape, which would have been fairly similar between the two observations.

The LMC data allowed a more thorough comparison with previous observations. Data for the LMC were gridded into a data cube of angular dimensions $8\times 8$ deg$^2$ and frequency extent 1~MHz (211 km~s$^{-1}$). The pixel size was 3 arcmin in RA and Dec, and 5 kHz (1.1 km~s$^{-1}$) in frequency (velocity). The gridding kernel was a circular Gaussian of FWHP width 8~arcmin, truncated at a radius of 9~arcmin. The small interior coverage gaps were filled using minimum-curvature 2D cubic interpolation (SciPy function {\tt interpolate.CloughTocher2DInterpolator}). The data cube was convolved with a further 8~arcmin Gaussian to give a final angular resolution of about 18 arcmin (the same as for the NGC 6744 cube).  {Finally, a small position shift (3 arcmin in RA; 6 arcmin in Dec)} was applied in order to align with previous LMC HI images from the multibeam \citep{2003MNRAS.339...87S}. 

The resultant channel images are shown in Figure~\ref{fig:LMC_channel_maps}, and the overall column density image is shown in Figure~\ref{fig:LMC_column_density}. The quality and dynamic range look excellent, with only low-level artifacts caused by the coverage gaps. The background noise level ($\sim 18$mK, in 5 kHz channels, corresponding to a column density rms of $3.6\times10^{16}$ cm$^{-2}$) looks uniform.  {For a noise-equivalent brightness temperature sensitivity of 25 K,} the theoretical Stokes $I$ noise for each beam is 28mK but, due to beam overlap, the mean spatial weight in the gridded cube is 2.5, leading to an expected image noise of $28/\sqrt 2.5=18$mK. We have not taken into account beam covariance,  {so this will be a lower limit}.

The features, positions and column densities show good agreement with previous multibeam images \citep{2003MNRAS.339...87S}. A more quantitative comparison using a pixel-pixel (T-T) plot is shown in Figure~\ref{fig:lmc-comparison}. To produce this, the multibeam image was resolution-matched by convolution with an 8 arcmin FWHP Gaussian, then regridded into the same RA-Dec-frequency coordinate system as the cryoPAF data. On the whole, the comparison is excellent. CryoPAF temperatures are approximately 0.97 of multibeam temperatures. This  {is} consistent with the updated S9 temperature \citep[83~K;][]{2005A&A...432...45B} used here, compared with the original low-resolution measurement \citep[85~K;][]{1973A&AS....8..505W} used for the multibeam calibration. 

 {However, Figure~\ref{fig:lmc-comparison} also shows that there a systematic offset between the datasets  in the sense that, despite the lower overall scale factor, the cryoPAF cube actually contains more signal at low temperatures. This offset is $\sim 0.1$~K for brightness temperatures $T_B<0.3$~K, rising to a maximum of $\sim 0.4$~K for $3 < T_B  < 10$~K. At velocities away from the LMC, the offset is only a few mK, which is a small fraction of the rms noise. What is origin of the offset? It appears to be real signal that has been over-subtracted from the multibeam cube due to poorer quality bandpass calibration and baseline removal (high-order polynomials were used in the spatial and frequency domains). Detailed inspection shows that the excess emission tends to be located around existing emission features, consistent with this interpretation. It is unlikely that these features are due to larger cryoPAF sidelobes. We were unable to make beam maps during the commissioning observations, so this remains to be verified. However, the sidelobes of the maxSNR offset beams measured during HI commissioning of the FLAG instrument were a level of $\sim 3$\% \citep{2021AJ....161..163P}, which is probably too low to provide an explanation.} 

 {With an estimated 20\% of pixels having emission in the temperature range 0.1 -- 1 K over the 211 km~s$^{-1}$ velocity range of the re-gridded multibeam data cube, this corresponds to a mean column density deficit of $8\times 10^{18}$ cm$^{-2}$. For an LMC distance of 50~kpc, the mass of the missing low-brightness-temperature HI is $\sim 3\times10^6$ M$_{\odot}$, or only about 0.6\% of its total HI mass. However, over all pixels, the  mean column density in the multibeam data is 9\% lower than for the cryoPAF measurements presented here. This implies that the true HI mass of the LMC is correspondingly higher than previously measured, so probably close to $5.2\times10^8$ M$_{\odot}$. In the context of diffuse atomic gas and `dark' molecular gas, accurate column density measurements of the ISM and in the outer fringes of galaxies are important for a clearer picture of gas accretion, star-formation efficiency and galaxy evolution in general \citep{2024ApJ...973...15W, 2020A&A...643A.141M}. The new data provides some important context when comparing interferometer data with older single-dish observations of the Milky Way and nearby galaxies.}

\section{Techniques for weak signal detection}
\label{sec:techniques}

The next step in scientific data analysis of cryoPAF data is to investigate more advanced data reduction techniques  {that} make better use of the high dimensionality of cryoPAF data, particularly for cosmological signals. As opposed to the LMC signals, which are bright, redshifted cosmological signals are faint and occur over a frequency range which includes much RFI, therefore need careful flagging and `de-noising'. Cosmological signals also need protection against signal loss, as illustrated in a minor manner in Section~\ref{sec:verification}. For this purpose, we utilise the data described above, but `inject' artificial signals so that we can verify the level of noise reduction achieved and the accuracy of signal recovery.

The zoom-band data for NGC~6744 is used as an example of fairly `clean' data in a band free from RFI, where it was straightforward to study the effects of signal capture and loss using different algorithms.  The 230 MHz wideband data of the LMC is used as an example of more typical `challenging' observational data in a band that contains RFI, and modulation in the sky continuum level due to thermal and non-thermal emission from the LMC. 
Example time-frequency waterfall plots of the central beam in the two sets of observations are shown in Figure~\ref{fig:waterfall1+2}.

\label{sec:methods}

\subsection{Singular value decomposition}

Singular value decomposition (SVD) is a form of multilinear regression which identifies the intrinsic dimensionality (or matrix rank) of a two-dimensional $m\times n$ array of measurements by elimination of linear dependencies. By sorting the amplitudes of the singular values (which are akin to the eigenvalues of principal component analysis), dimensionality can be further reduced by including only the most significant sources of variance. 

In the context of image compression, this allows features which are common in datasets to be retained, and uncorrelated features  {(such as random noise)} to be removed. However, in radio astronomy, the strongest and most common features include wideband RFI bursts, constant narrowband RFI, changes in the sky continuum level, changes in receiver gain,  {and even correlations due to beam overlap and the spectral point source response. These are noise, non-signal or redundant features which we usually desire to {\it remove}, thereby leaving signals of interest}. Compression of data is therefore usually minor, but removal of artifacts can be significant. 

In spectral-line radio astronomy, common dimension pairs are frequency and time, or frequency and position.
If $\mathbf{M}$ is the real-valued $m\times n$ matrix representing these measurements, its full decomposition is given by:

\begin{equation}
\mathbf{M} = \mathbf{U} \mathbf{S} \mathbf{V}^{\rm T} ,
\label{eq:svd}
\end{equation}
where $\mathbf{S}$ is the $m\times n$ matrix of singular values and $\mathbf{U}$ and  $\mathbf{V}$ are orthogonal $m\times m$ and $n\times n$ matrices, respectively. The truncated approximation to $\mathbf{M}$ is then given by:
\begin{equation}
\mathbf{M} \approx \mathbf{U}_r \mathbf{S}_r \mathbf{V}^{\rm T}_r ,
\label{eq:truncatedsvd}
\end{equation}
where $\mathbf{S}_r$ is a compact $r\times r$ matrix of the largest singular values and $\mathbf{U}_r$ and  $\mathbf{V}_r$ are the $m\times r$ and $n\times r$ matrices of column vectors corresponding to the largest singular values. After sorting by significance, these matrices are unique.

Although straightforward to compute, the above SVD and truncated SVD forms have several disadvantages, including inability to deal with non-linear dependencies, inability to deal with missing data, and sensitivity to outliers  {(see also Section~\ref{sec:intro})}. In the current context, non-linearities are less important, but can be dealt with using unsupervised machine learning techniques. Missing data and outliers tend to be related -- outliers are either present in the data, or they are flagged as `bad' by the observer. Outlier rejection can be dealt with by iterative SVD techniques (see Section~\ref{sec:results}), or by more sophisticated L1 or robust kernel methods \citep{1467342, 9747522, 9919215, pmlr-v238-han24a}.

\subsection{Higher-order singular value decomposition}

Higher-order or tensor SVD  is a generalisation of the two-dimensional SVD method which reduces the number of components required to describe each of the dimensions (modes) in a three, or higher, dimensional dataset. For  {a} tensor of order $N$, this can be achieved by unfolding, or flattening, the tensor along $N-2$ of its dimensions and applying the above 2D, or matrix-based SVD technique. However,  {true} higher-order techniques maintain tensor order. Therefore, these techniques have found use in multidimensional event detection, data modelling, interpolation of missing data, and data compression. 

Similar to 2D SVD, different methods of higher-order decomposition have been proposed \citep{Tucker_1966, doi:10.1137/110837711, 7782758, AHMADIASL2024168}, and different methods are available to deal with outliers and missing data \citep{8646385}. The Canonical Polyadic or CANDECOMP/PARAFAC (CP) SVD approximation of a tensor of order-$N$ can be written in outer product notation \citep{gillet:hal-03892165} as:

\begin{equation}
\mathbf{M} \approx \sum_{\ell=1}^n \mathbf{U}^{(1)}_{\ell} \circ \mathbf{U}^{(2)}_{\ell} \circ  ... \circ \mathbf{U}^{(N)}_{\ell} ,
\label{eq:cp-svd}
\end{equation}
where $n$ is the CP rank, or number of singular values required for the approximation to be valid \citep{2009SIAMR..51..455K}. 

In the the alternative method of Tucker decomposition \citep{Tucker_1966}, the approximation is given by the result of mode-$k$ products:
\begin{equation}
\mathbf{M} \approx \mathbf{S} \times_1 \mathbf{U}^{(1)} \times_2 \mathbf{U}^{(2)} \times_3 ... \times_N \mathbf{U}^{(N)} ,
\label{eq:tucker-svd}
\end{equation}
where $\mathbf{S}$ is a core, or compact, tensor of reduced rank $\mathbf{r} = (r_1, r_2, ..., r_N)$ and the $\mathbf{U}^{(k)}$ are $n_k \times r_k$ orthonormal factor matrices. 

In yet another alternative (not implemented here), the t-SVD tensor completion method, the reduced 3D SVD is given by the t-product \citep{7782758}:
\begin{equation}
\mathbf{M} \approx \mathbf{U} * \mathbf{S} * \mathbf{V}^{\rm T} ,
\label{eq:reducedt-svd}
\end{equation}
where the tensor $\mathbf{S}$ has dimensionality ($n, n, n_3$), with $n$ being the rank-$n$ compression of $\mathbf{M}$.

\subsection{Data dimensionality}
\label{sec:dimensionality}

 {
The observational datasets already described each consists of 72 beams, with each beam having a large number of spectra, each taken at a different time, and RA/Dec sky position. These datasets are therefore 4-dimensional, but irregularly sampled, which is inconvenient for the present purposes. We are only interested in the noise and RFI characteristics of the dataset, as we will be creating our own `signals' by injection. 
}

 {
Therefore, we have `re-folded' the data such that time is identified with first spatial dimension (which it is, even though sampling is irregular), and beam number is identified with the second spatial dimension (which it also is, but again irregular). Frequency remains the third dimension -- this remains regularly sampled and monotonic. The advantage of this re-folding is that the resultant dataset is a fully-sampled (tensor) array, to which we can inject signal as if it was a three-dimensional cube of the sky. The sampling remains monotonic in all dimensions, so the spatial correlations are retained, at least prior to SVD.
}

 {
In practice, in a targetted observation with much more observing time, a 4-dimensional array would be preferred. Ideally this would be 72 data cubes, each using the same RA-Dec-frequency grid. 
Other dimensions such as polarisation and pulsar bin can be added by further increasing tensor order (or re-shaping). SVD can also be applied to complex cross-polarisation or interferometric data by replacing $\mathbf{V}^{\rm T}$ in Equations \ref{eq:svd} and \ref{eq:truncatedsvd} with the corresponding conjugate transpose $\mathbf{V}^{\rm H}$.
}

\subsection{Algorithms}

We have chosen to compare the following SVD algorithms, representing a mix of 2D (1--3), pseudo-3D (4) and 3D tensor (5--6) methods, with and without some form of robustness using an L1 loss function, trimming, or censorship:
\begin{enumerate}
\item SVD: we use standard SVD truncation techniques using the Python {\tt numpy.linalg.svd} implementation.
\item L1SVD: to minimise the effect of outliers, which are common in the presence of RFI or strong astronomical signals, we use L1-norm error minimisation low-rank estimation algorithm of \cite{1467342} as implemented by \cite{9919215}.
\item CSVD: as L1 techniques are slow, we also implement a clipped version of standard SVD, which iteratively trims outliers above and below the 99.5 and 0.5 percentiles to the data values at those percentiles, respectively (Winsorisation).
\item CSVDstack: this is a common technique, in which the third (or higher) dimension is unfolded into a lower dimension, for example by reshaping the data array. As for CSVD, we have implemented iterative trimming, for fast, reliable and robust results.
\item TuckerSVD: Tucker tensor decomposition \citep{Tucker_1966, 2009SIAMR..51..455K} is a true higher-order technique which approximates the input data by a low-rank core tensor. For convenience, we treat the reduced rank for each dimension equally ($r_1=r_2=r_3$), but do not allow $r$ to exceed the size of the smallest dimension. We use the Python function {\tt tensorly.decomposition.tucker} \citep{JMLR:v20:18-277} which permits outliers to be masked. We iteratively censor data above and below the 99.5 and 0.5 percentiles, respectively.
\item CPSVD: CP decomposition results in a set of parallel factor matrices of low rank, whose outer products are used to define an approximation to the input tensor. We use the Python function {\tt tensorly.decomposition.parafac} \citep{2009SIAMR..51..455K} which again permits outliers to be iteratively censored.
\end{enumerate}

\begin{figure*}[t]
    \centering
    \includegraphics[width=0.49\textwidth]{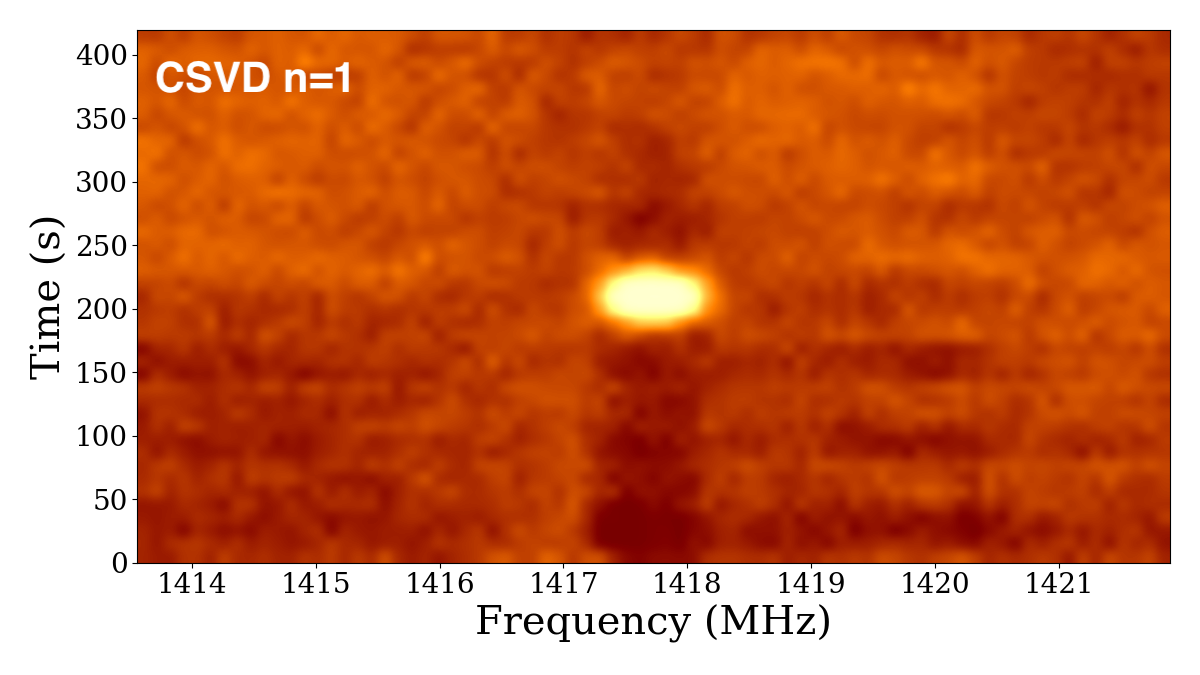}
    \includegraphics[width=0.49\textwidth]{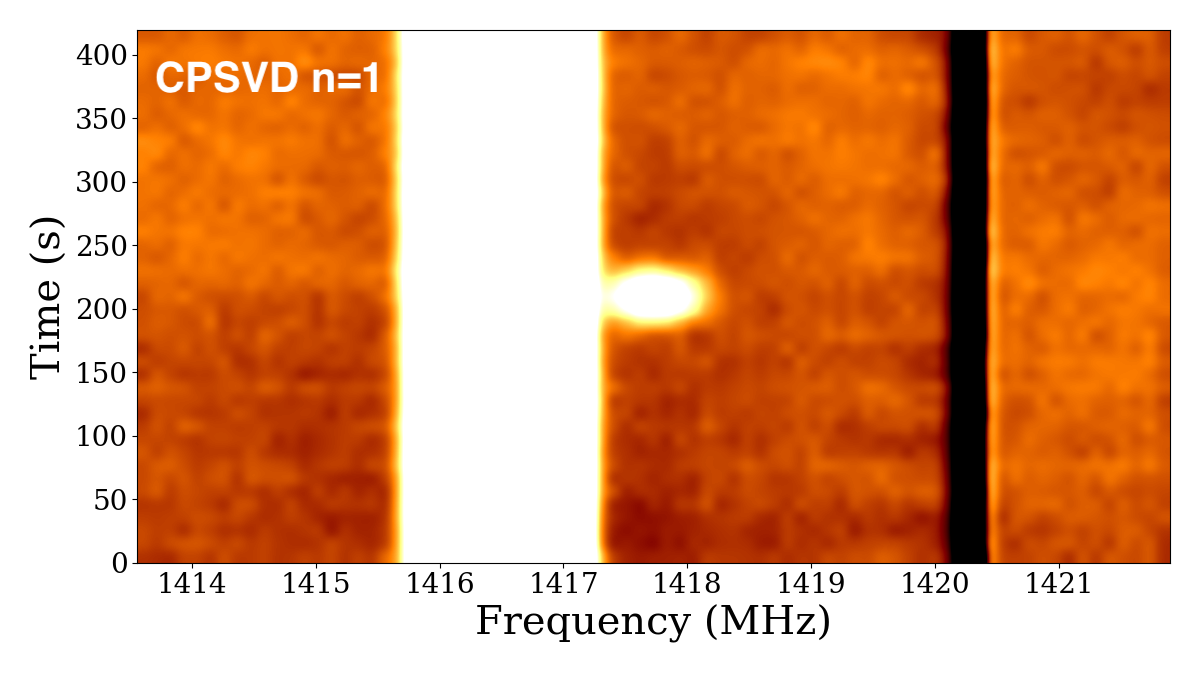}
    \includegraphics[width=0.49\textwidth]{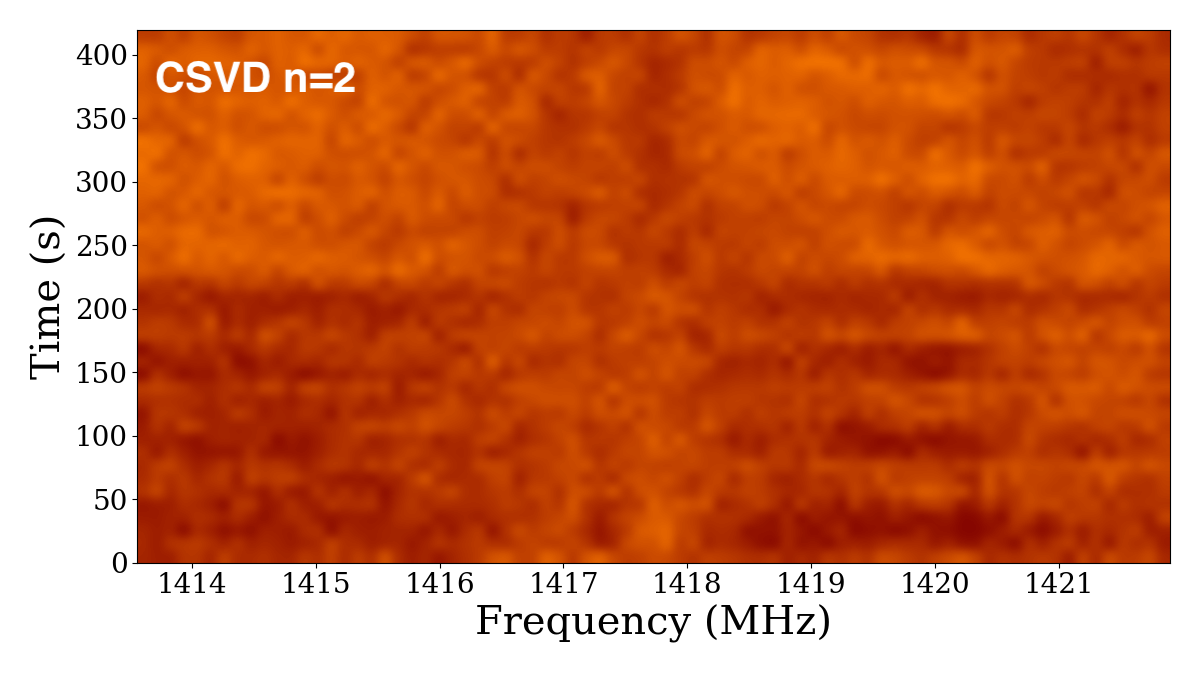}
    \includegraphics[width=0.49\textwidth]{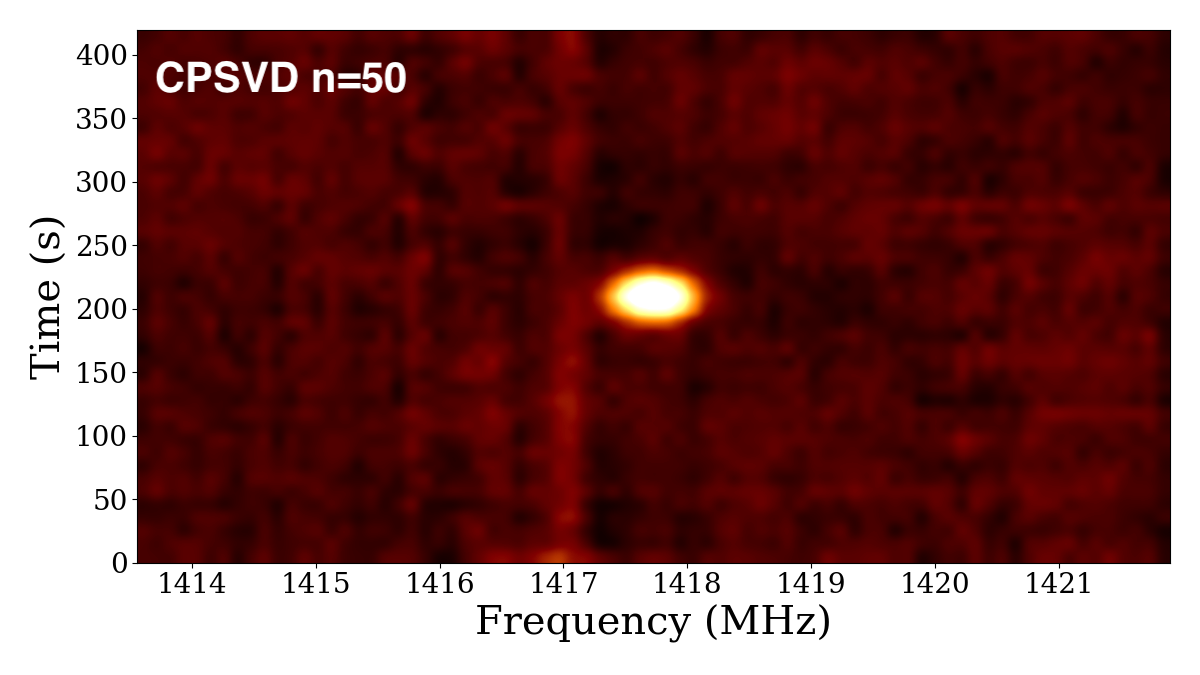}
    \caption{Frequency--time waterfall plots of beam 36 in the NGC 6744 dataset after application of ({\it left column}) a clipped two-dimensional SVD (CSVD) and ({\it right column}) a censored three dimensional SVD (CPSVD). Prior to SVD, a 1~Jy compact source was injected at the central frequency  and time. The best (in terms of S/N in Figure~\ref{fig:signal_loss}) CSVD results ($n=1$ and $n=2$) and CPSVD results ($n=1$ and $n=50$) are shown. Application of $n=1$ CSVD ({\it top left}) has completely removed the (non-varying) flux from NGC 6744 and the Galaxy, but  {retains the injected signal, albeit with significant loss of flux density and a negative sidelobe in the time dimension}. Application of $n=2$ CSVD ({\it bottom left}) removes flux even from the compact injected source. $n=1$ CPSVD ({\it top right}) retains NGC 6744 flux, but does not flatten the background as well. $n=50$ CPSVD ({\it bottom right}) removes much of the NGC 6744 flux but retains most  {(76\%)} of the injected flux. The intensity range is $-$0.1 to 0.6 Jy for all plots. The `pre-SVD' plot is show in the left-hand panel of Figure~\ref{fig:waterfall1+2}.}
    \label{fig:flattening}
\end{figure*}

\begin{figure*}[t]
    \centering
    \includegraphics[width=0.43\textwidth]{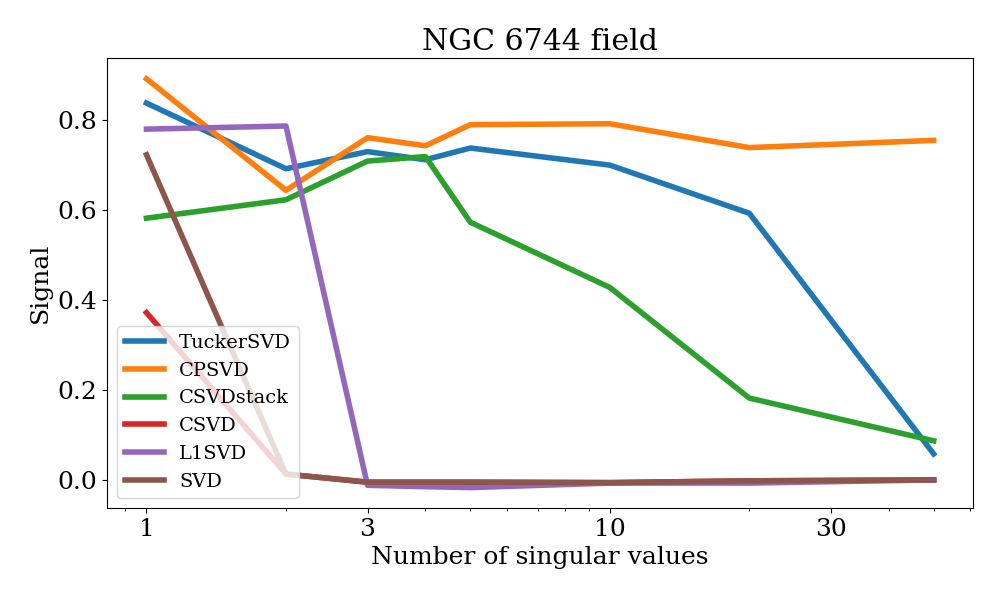}
    \includegraphics[width=0.43\textwidth]{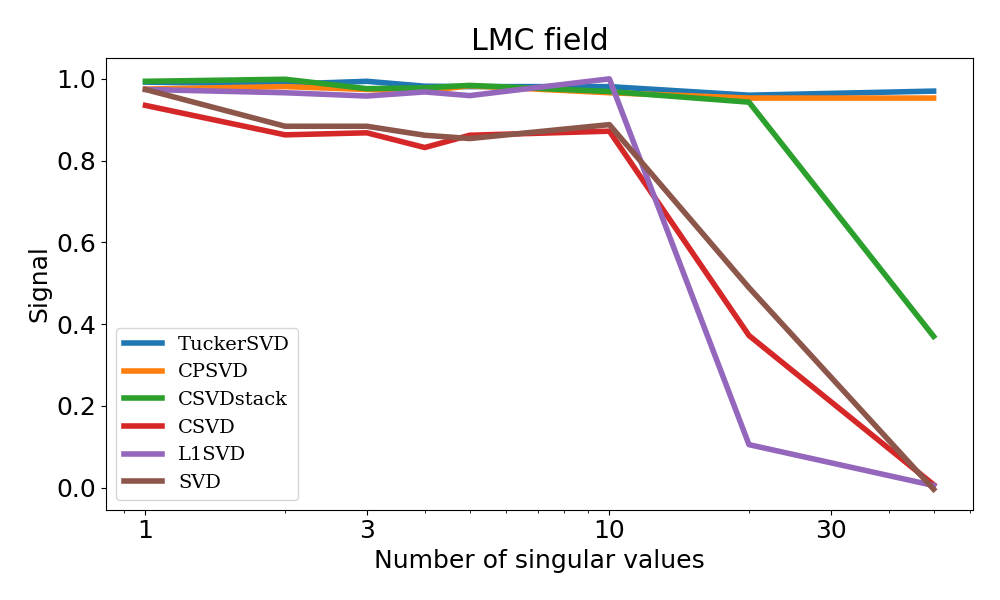}
    \includegraphics[width=0.43\textwidth]{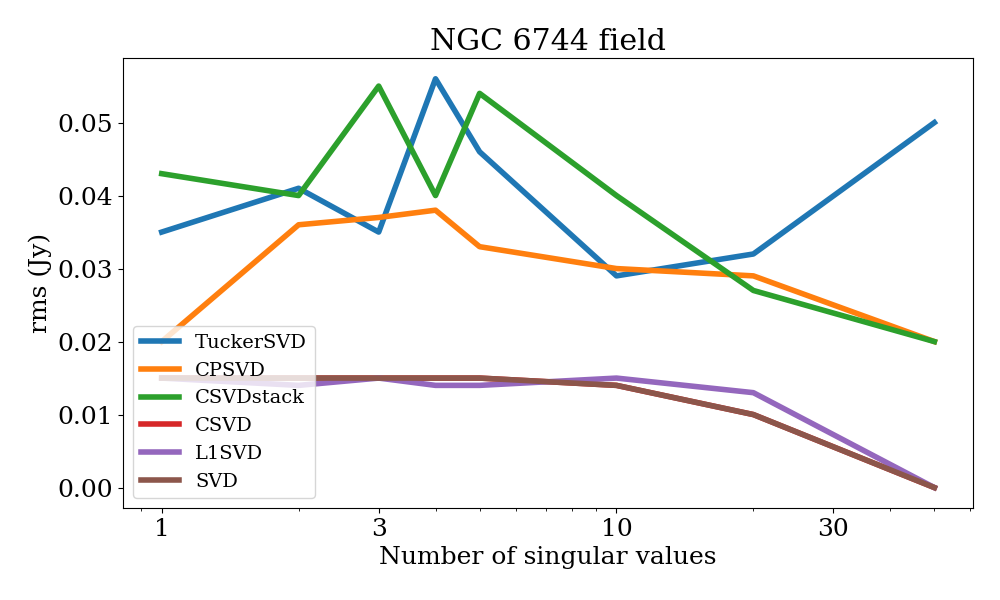}
    \includegraphics[width=0.43\textwidth]{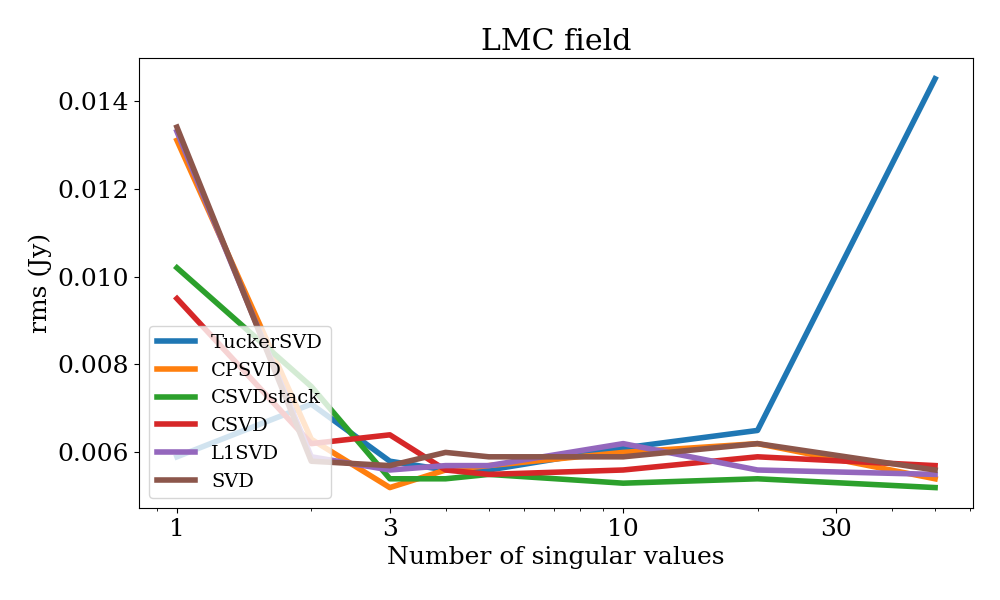}
    \includegraphics[width=0.43\textwidth]{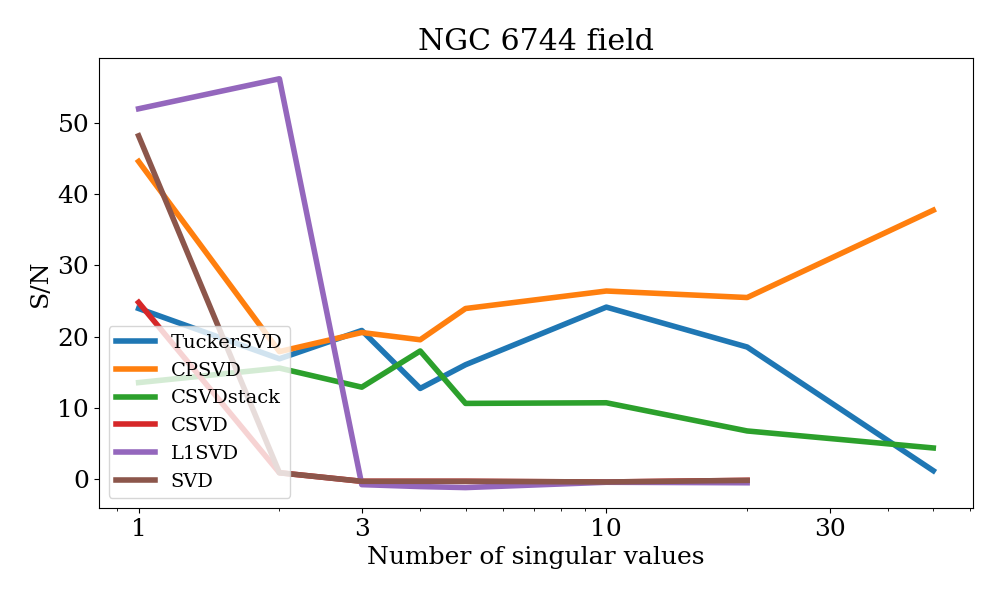}
    \includegraphics[width=0.43\textwidth]{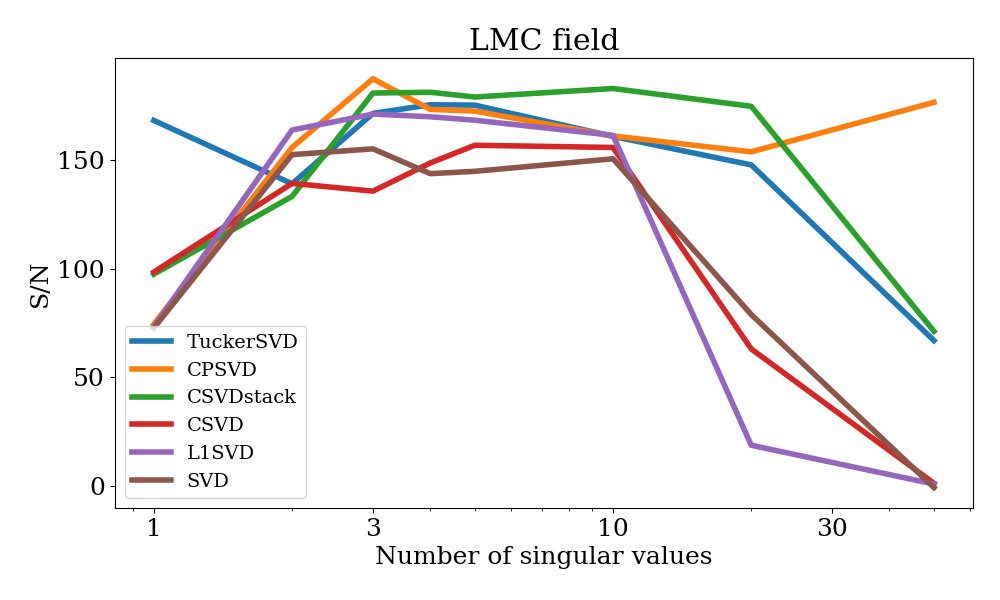}
    \caption{ {The {\it top row} shows the recovered signal (i.e.\ the ratio of measured signal to the injected input  signal) for a compact source in {\it (left)} the `clean' NGC 6744 field and {\it (right)} the `challenging' LMC field, as a function of the number of singular values removed}. The middle row shows the corresponding rms without any signal injection. The bottom row shows the corresponding signal-to-noise ratio S/N. In the case of the 2D SVD, CSVD and L1SVD methods, the number of singular values plotted is the number removed for a single beam. In the case of the CSVDstack method, it is the total number of values removed for the whole 3D 72-beam data cube. In the case of the CPSVD method, it is the number of outer products removed (the rank of the factor matrices). In the case of the TuckerSVD decomposition, it is the rank of each dimension in the core 3D tensor.}
    \label{fig:signal_loss}
\end{figure*}

\subsection{Pre-conditioning}
\label{sec:precon}

Pre-conditioning can usefully speed convergence of SVD and tensor SVD, reduce the effect of outliers, make comparison between different methods 
fairer, and allow physically-motivated normalisation and correction of various instrumental responses prior to unsupervised dimensionality reduction. In the following analysis, we have applied calibrations as described in Section~\ref{sec:calibration}. We also remove gradients in frequency and time to account for spectral response and continuum sources.
These techniques are normal in multibeam single-dish spectral-line radio astronomy, and descriptions are available for data from the Murriyang multibeam \citep{2001MNRAS.322..486B, 2014PASA...31....7C, 2016AJ....151...52S}, Arecibo ALFALFA \citep{2018ApJ...861...49H} and the FAST multibeam \citep{2022PASA...39...19X, 2023ApJ...944..102W} instruments.
 {A second consideration is that pre-conditioning permits the injection of fake sources without having to apply `inverse calibration' to the injected signal, so correct signal strength and S/N ratio is maintained for the injected signal.}

In summary, the following pre-conditioning steps were applied:

\begin{enumerate}
    \item Gain calibration for all beams, based on a prior observation of a brightness temperature or flux density calibrator.
    \item Bandpass calibration, based on an `off-source' observation at a different sky position.
    \item Subtraction of the median spectral bandpass (only for the LMC field). 
    \item Subtraction of the median power as a function of time (only for the LMC field).
    \item Spectral smoothing and sub-sampling to a resolution commensurate with the resolution of structures to be detected. We use  {0.11 MHz (23 km s$^{-1}$) and 0.89 MHz (190 km~s$^{-1}$ at $z=0$) boxcar smooths} 
     for the NGC 6744 zoom band and the  LMC wideband data, respectively. 
    \item  {For the extended source tests (discussed below), using the LMC wideband data, a further temporal smoothing of 30~s is applied. Frequencies above 1416 MHz ($z<0.003$) are also removed.}
    \item Normalisation of the data to zero median, unit median absolute deviation.
\end{enumerate}

 {These steps result in cube/tensor dimensions of (41, 72, 77) and  (388, 72, 263) for the NGC 6744 and LMC data, respectively. For the extended source tests in the LMC field, the tensor dimensions are (130, 72, 133). The dimension labels are time, beam and frequency, respectively (Section~\ref{sec:dimensionality}).}

\subsection{Signal injection}
\label{sec:signal}
To quantify the efficacy of the various SVD methodologies in removing instrumental, foreground, RFI and other artefacts, we inject artificial signals into the two datasets. These artificial signals are injected immediately before the normalisation stage (\#7) above. Two types of sources are injected:

\begin{enumerate}
    \item A compact source occupying $\sim 3$\% of the extent of the data in each dimension (i.e. frequency, time/position, and beam number), and at an approximately central position in each dimension. To avoid overlap, the source was placed at a slightly higher frequency than the HI emission of NGC 6744 and a lower frequency to the HI emission of the Galaxy/LMC.

    \item An extended HI intensity map derived from the catalogue of star-forming galaxies in the GAMA G23 redshift catalogue \citep{2022MNRAS.513..439D} in the redshift range $0.01 < z < 0.24$, spatially convolved by a Gaussian of FWHM $\approx 6$ cMpc and spectrally convolved with a Gaussian of width $\Delta z \approx 10^{-3}$. Only galaxies with H$\alpha$ equivalent widths significantly ($3\sigma$) above zero and in the main G23 region were included \citep{2017MNRAS.465.2671G}. We correct for sample selection using the `SF complete' function of \citet{2018MNRAS.479.1433G}.
\end{enumerate}

The amplitude of the compact source was set at 1 Jy for the NGC 6744 dataset and 0.1 Jy for the lower frequency resolution (and therefore lower theoretical noise) LMC wideband dataset. This corresponds to approximately 4\% and 0.4\% of the system equivalent flux density, respectively.

The average amplitude for the extended `source' was normalised to the HI sky brightness temperature \citep[e.g.][]{2013MNRAS.434.1239B}:
\begin{equation}
T_B = 63 h \left(\frac{\Omega_{\rm HI}(z)}{3.5\times10^{-4}}\right) \frac{(1+z)^2}{\sqrt{\Omega_{\rm m} (1+z)^3 + \Omega_{\Lambda}}}   ~~ \mu \mathrm{K} ,
\label{eq:TB}
\end{equation}
assuming $\Omega_{\rm HI}(z) \approx 3.5\times 10^{-4} (1+z)^{0.63}$ for $z<5$ \citep{2023MNRAS.518.4646R}, Planck 2018 cosmological parameters $(h, \Omega_{\rm m}) = (0.674, 0.315)$ \citep{2020A&A...641A...6P}, and $\Omega_{\Lambda}=1-\Omega_{\rm m}$.
To better understand signal loss, noise reduction and the efficacy of various SVD methods, the LMC data  {were} artificially scaled to a regime where the detection of the injected G23 signal was `challenging' (signal-to-noise ratio less than unity). 

\begin{figure*}[t]
    \centering
    \includegraphics[width=0.45\textwidth]{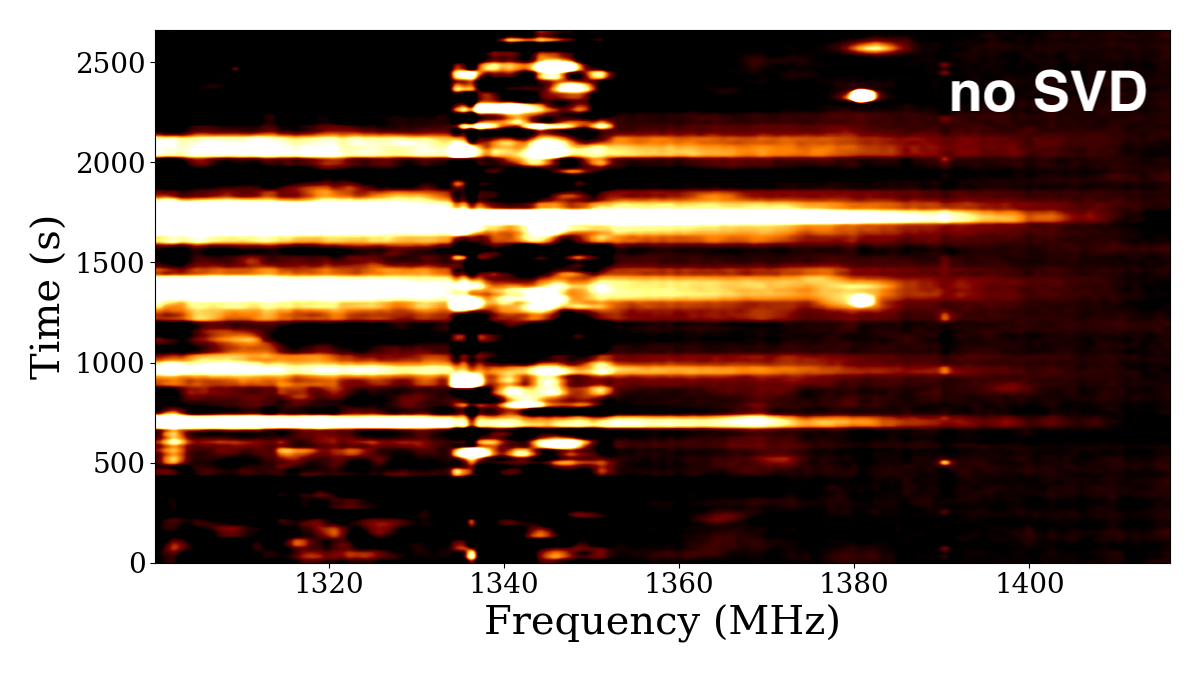}
    \includegraphics[width=0.45\textwidth]{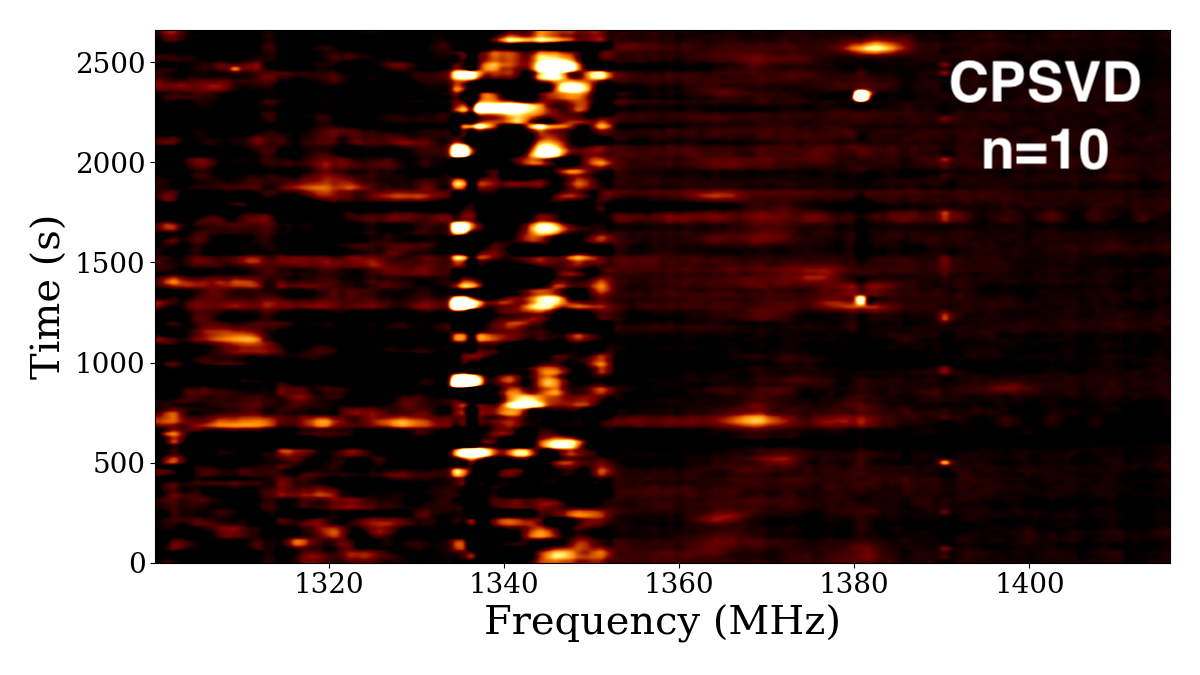}
    \includegraphics[width=0.45\textwidth]{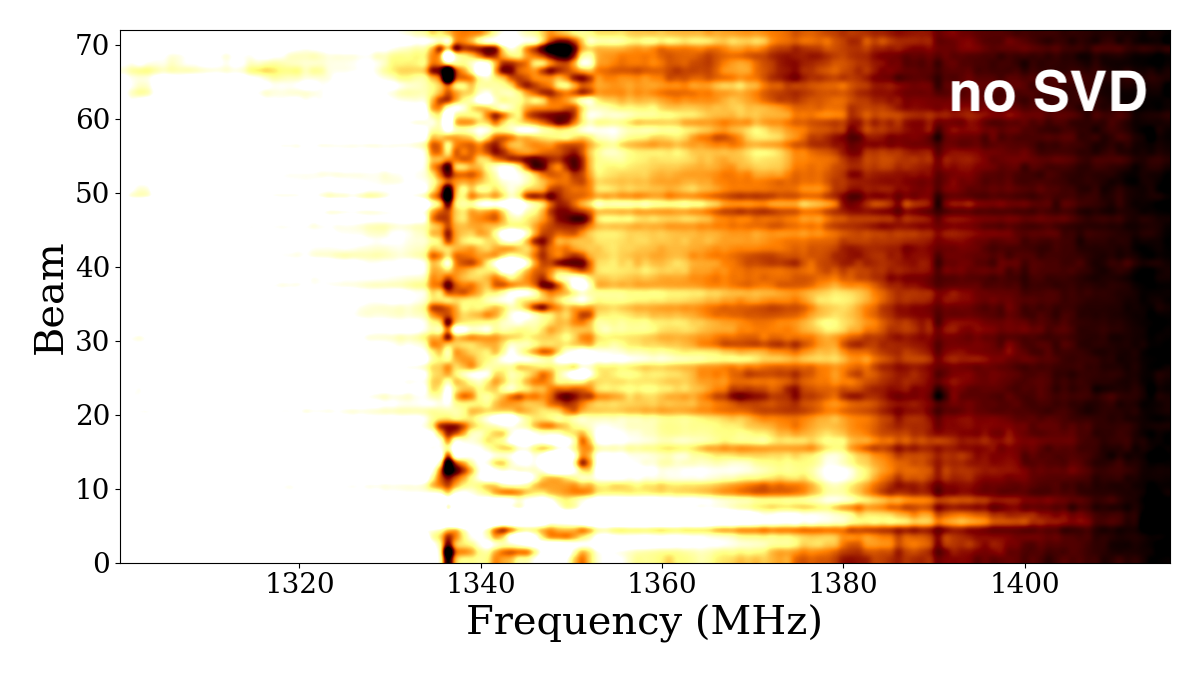}
    \includegraphics[width=0.45\textwidth]{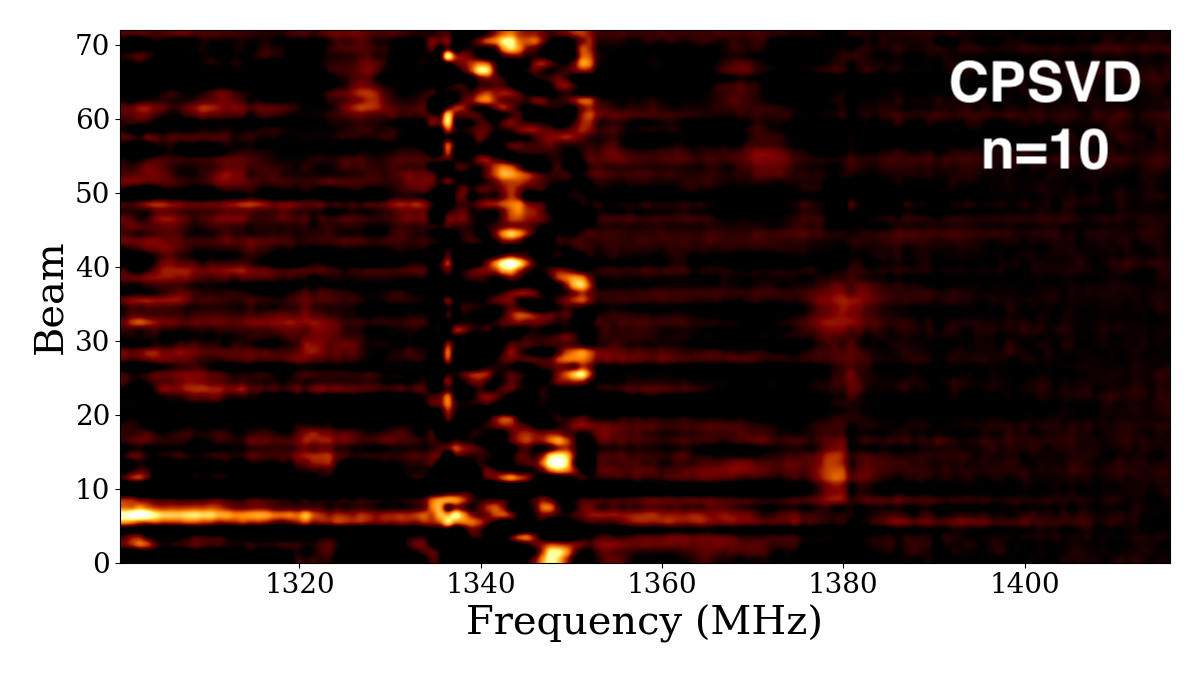}
    \includegraphics[width=0.45\textwidth]{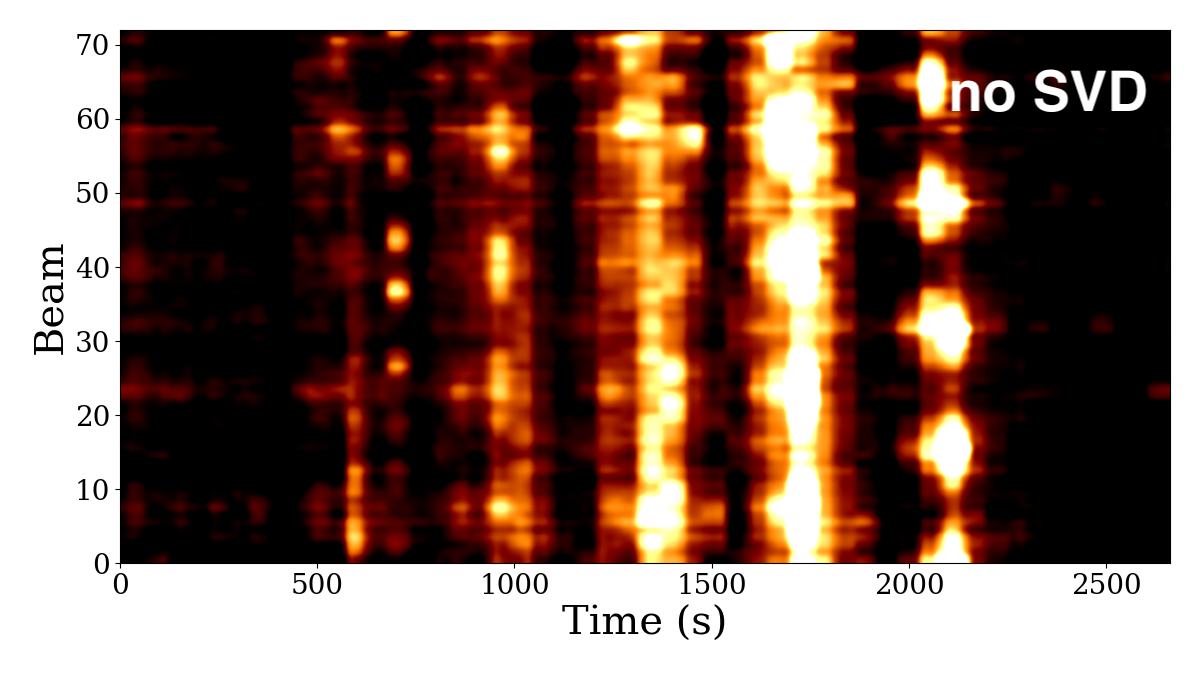}
    \includegraphics[width=0.45\textwidth]{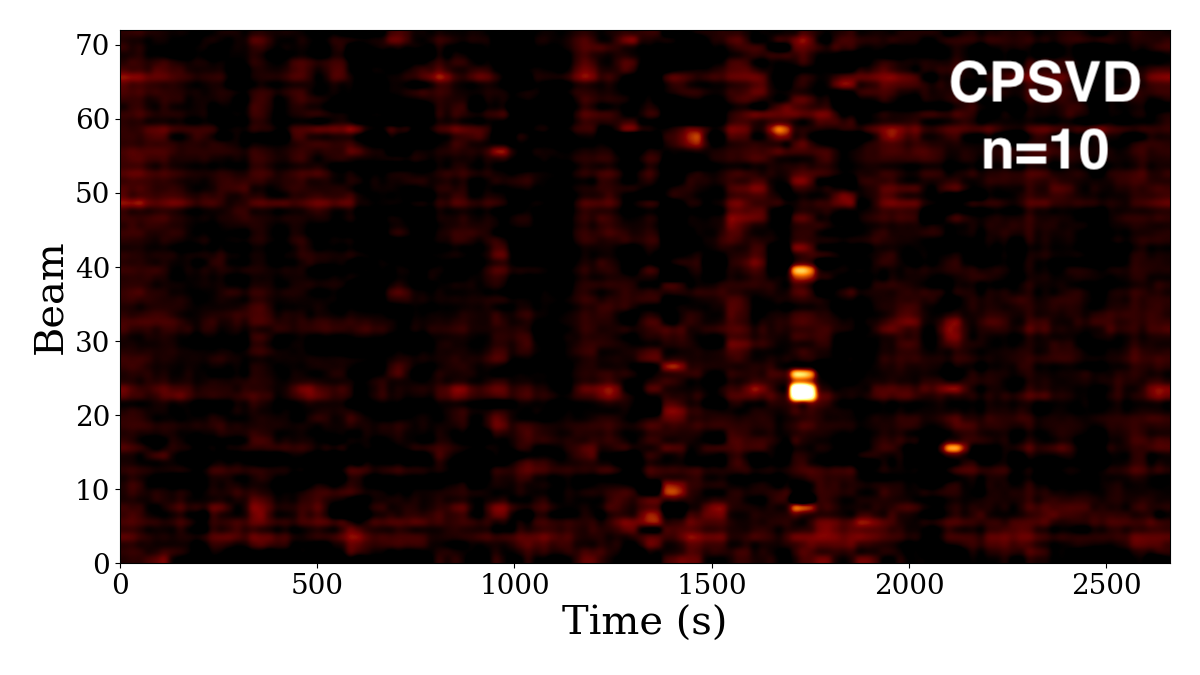}
    \caption{Waterfall plots of example 2D sections of the LMC dataset after addition of the G23 HI intensity map, scaled by a factor of $10^2$. From {\it top} to {\it bottom}, the three sections are: frequency-time; frequency-beam; and time-beam (the time dimension is also a spatial dimension due to the changing telescope position). The {\it left} column represents the data after step 5 of pre-conditioning in Section~\ref{sec:precon} -- i.e.\ only basic temperature calibration and smoothing. The {\it right} column is an example of the same data after SVD application -- in this case Canonical Polyadic SVD (CPSVD) with 10 singular values removed. The band of emission at $1343\pm9$ MHz is RFI, also faintly seen in the right-hand panel of Figure~\ref{fig:waterfall1+2}, prior to pre-conditioning. The stripes along the time axis in the pre-SVD plots in the left column are receiver gain variations. The faint `blobs', seen mainly in the CPSVD plots, are the high peaks of the G23 intensity map. The slices are all taken at the mid-point of the hidden dimension (e.g.\ the upper waterfalls are for beam 36 out of 72). The intensity range is $-20$ to 200mK for all plots.}
    \label{fig:lmcflattening}
\end{figure*}

\begin{figure*}[t]
    \centering
    \includegraphics[width=0.4\textwidth]{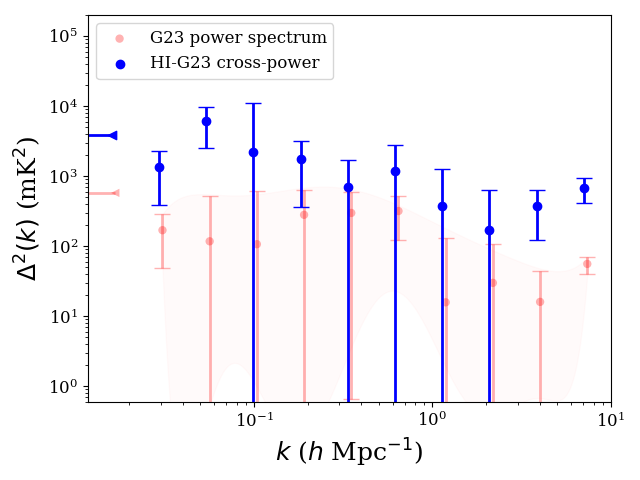}
    \includegraphics[width=0.4\textwidth]{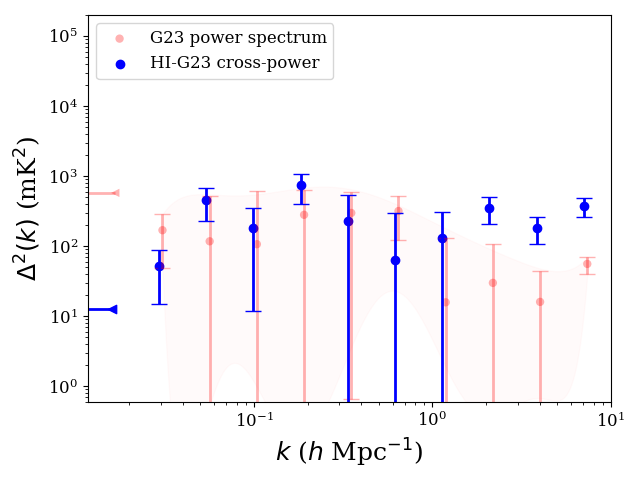}
    \caption{The {\it left} panel shows the power spectrum (red points) of the G23 intensity map. It also shows the cross-power spectrum (blue points) of the intensity map with the LMC dataset (which includes the co-added intensity map, scaled by $10^2$). The {\it right} panel shows the same intensity map power spectrum, but shows the cross-power spectrum (blue points) after CPSVD with $n=10$. The HI intensity field is much closer to the expected result after CPSVD application. Note that the error bars for the G23 power spectrum (red points) are reflective of sample variance due to the limited volume -- since this is a model, the actual errors are negligible. The red shaded region is a smoothed approximation to the error bars.}
    \label{fig:PSexamples}
\end{figure*}

\begin{figure*}[t]
    \centering
    \includegraphics[width=0.33\textwidth]{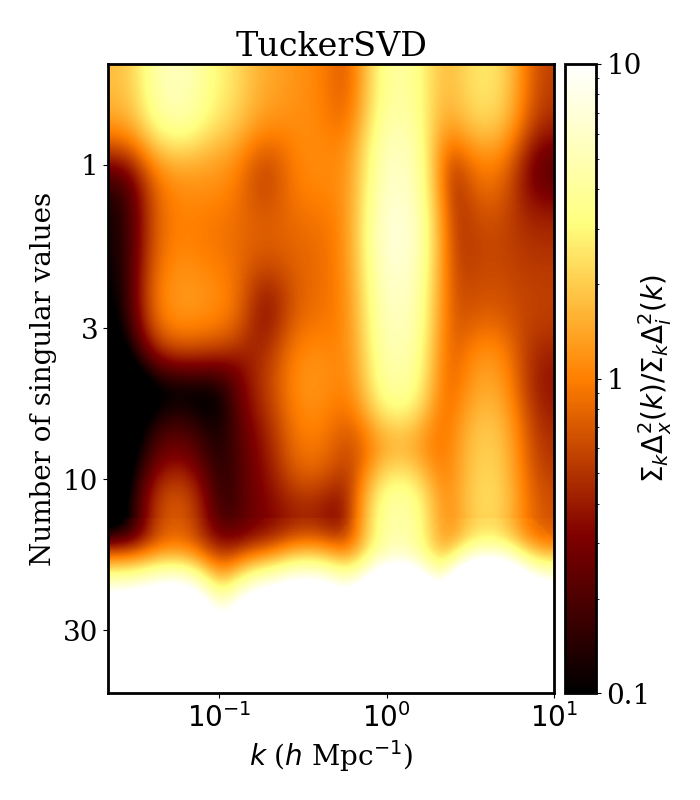}
    \includegraphics[width=0.33\textwidth]{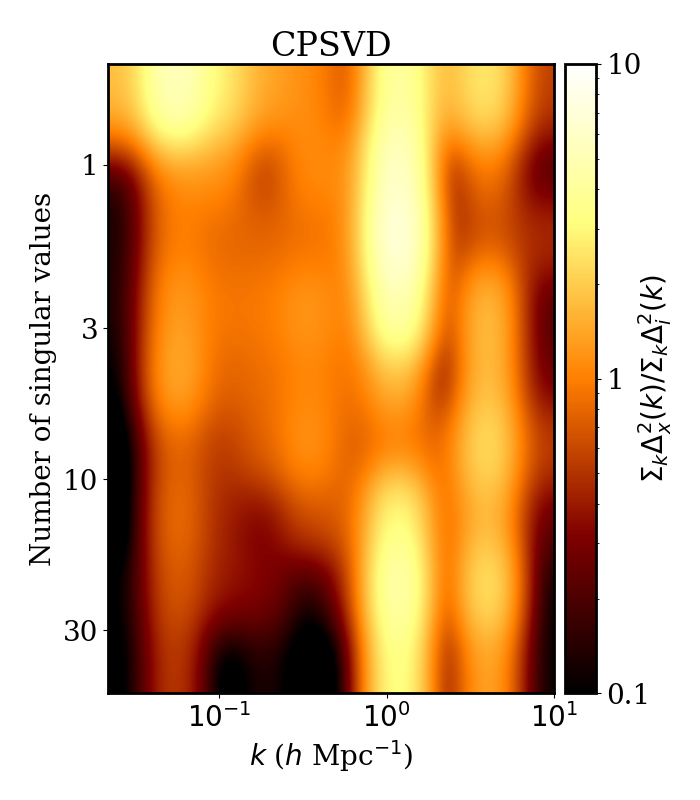}
    \includegraphics[width=0.33\textwidth]{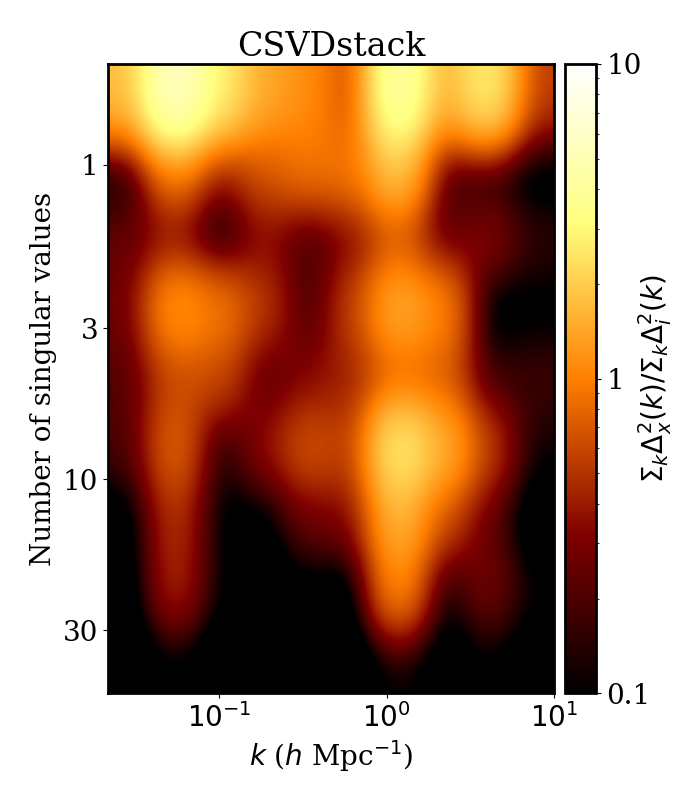}
    \includegraphics[width=0.33\textwidth]{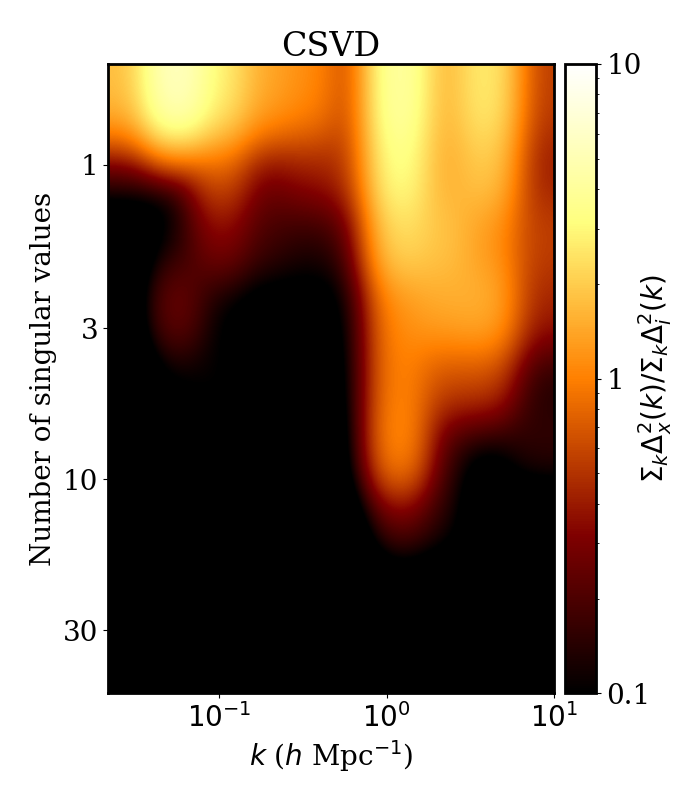}
    \includegraphics[width=0.33\textwidth]{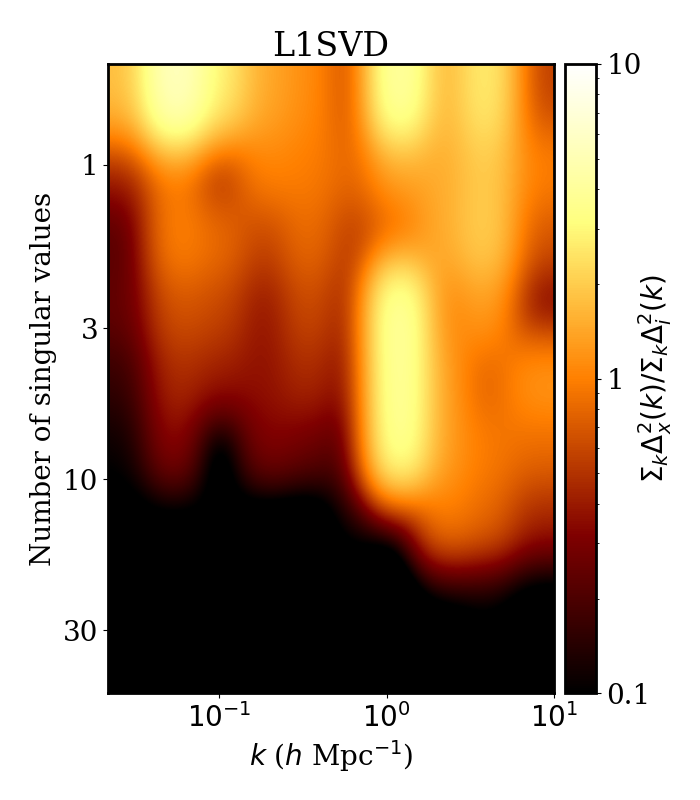}
    \includegraphics[width=0.33\textwidth]{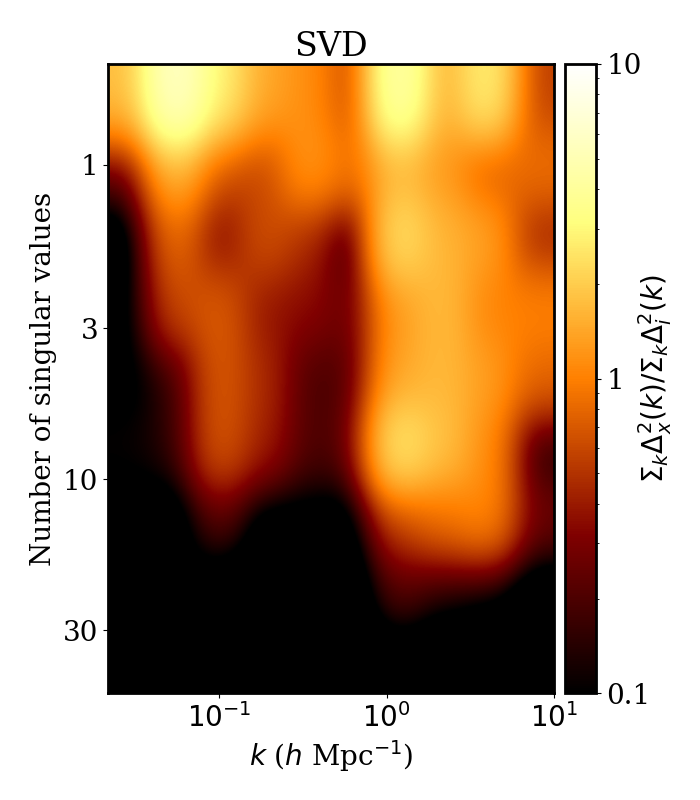}
    \caption{The colour represents the ratio of the amplitude of the cross-power spectrum to the amplitude of the G23 intensity map power spectrum scaled by $10^3$ as a function of both wavenumber and the number of singular values subtracted from the data. High values of the ratio (white) indicate noise contamination. Values of the ratio near unity (orange/red) mean that the intensity field power spectrum is unbiased; values less than unity (the darkest colours) indicate severe signal loss. Generally, the tensor SVD methods ({\it top} row) perform reasonably well at $n\approx5$. The 2D methods ({\it bottom} row) have similar performance to tensor SVD methods at large $k$ values, but tend to over-subtract signal low $k$ for $n>3$. As shown previously for compact sources, the Tucker approximation fails for $n>20$.}
    \label{fig:PSnsvdgallery}
\end{figure*}

\begin{figure}[t]
    \centering
    \includegraphics[width=0.9\textwidth]{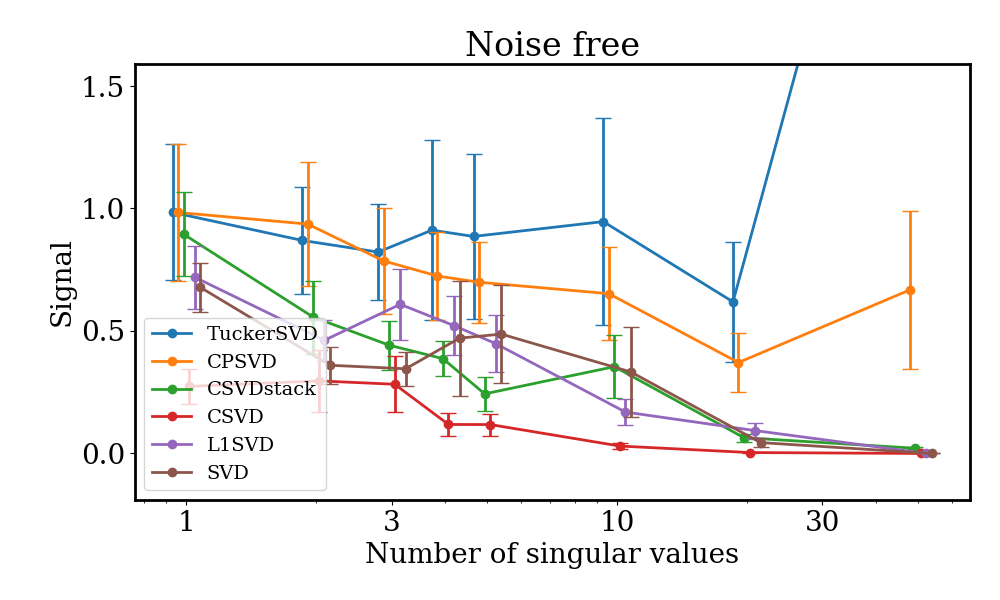}
    \includegraphics[width=0.9\textwidth]{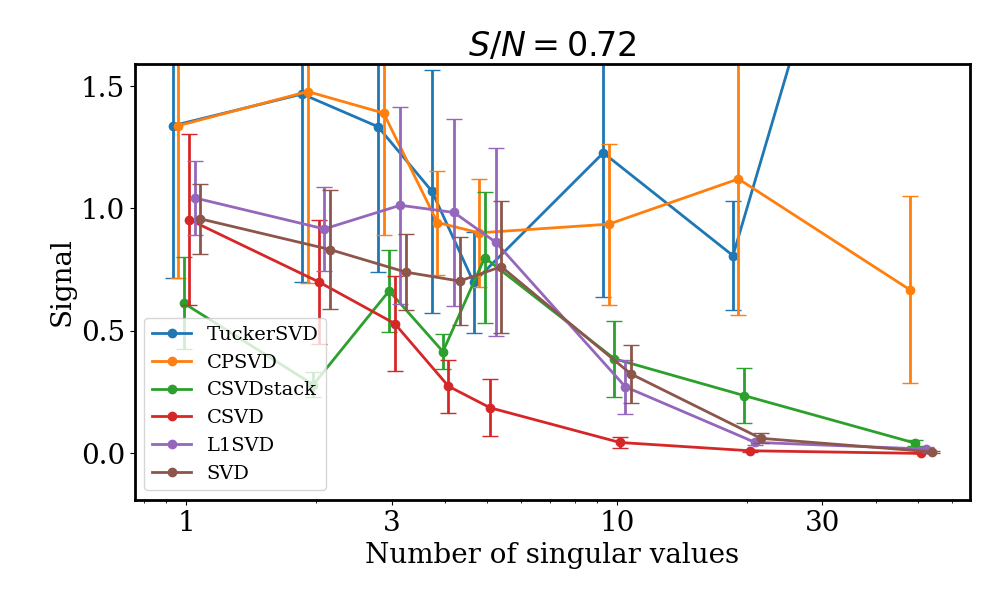}
    \includegraphics[width=0.9\textwidth]{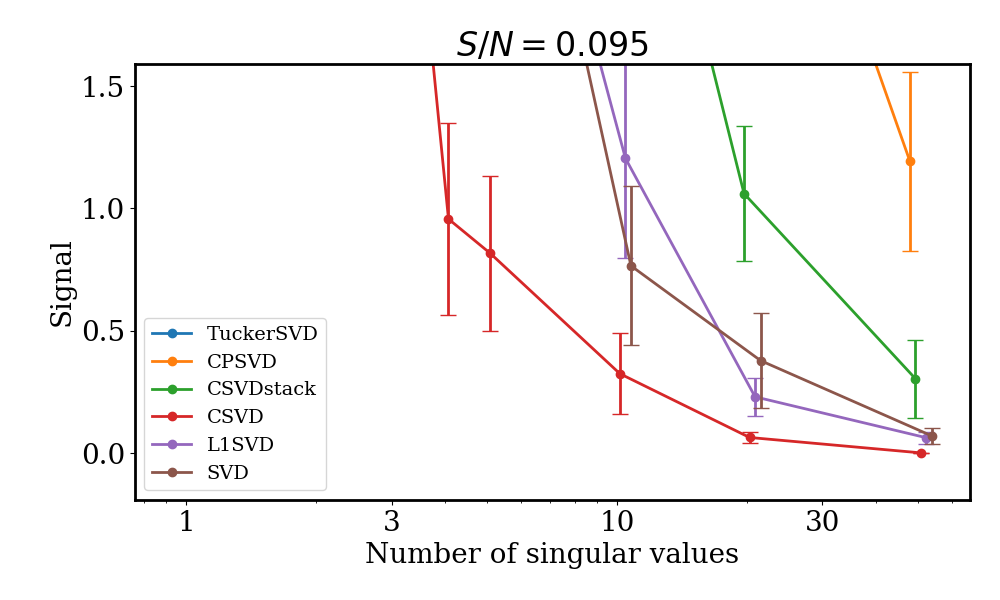}
    \caption{ {Normalised intensity map signal as a function of the number of singular values removed from the LMC/G23 dataset for the six SVD methods. The signal is defined as the ratio of the cross-power spectrum to the original intensity map power spectrum, averaged over all $k$ (i.e.\ 1.0 means there is no signal loss, and 0.0 means there is no remaining signal).}}
    \label{fig:IM_signal_loss}
\end{figure}

\begin{figure}[t]
    \centering
    \includegraphics[width=0.9\textwidth]{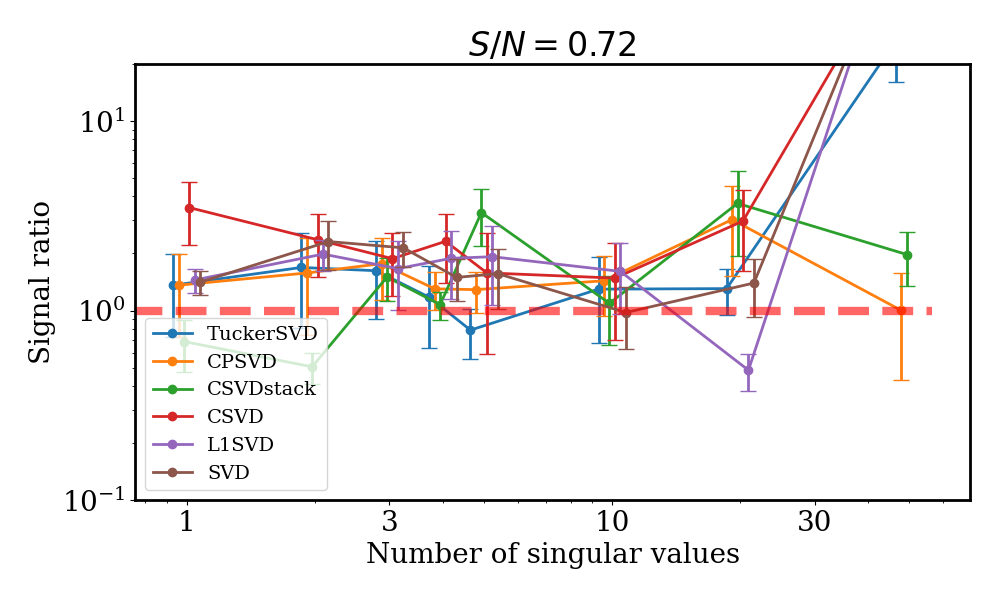}
    \includegraphics[width=0.9\textwidth]{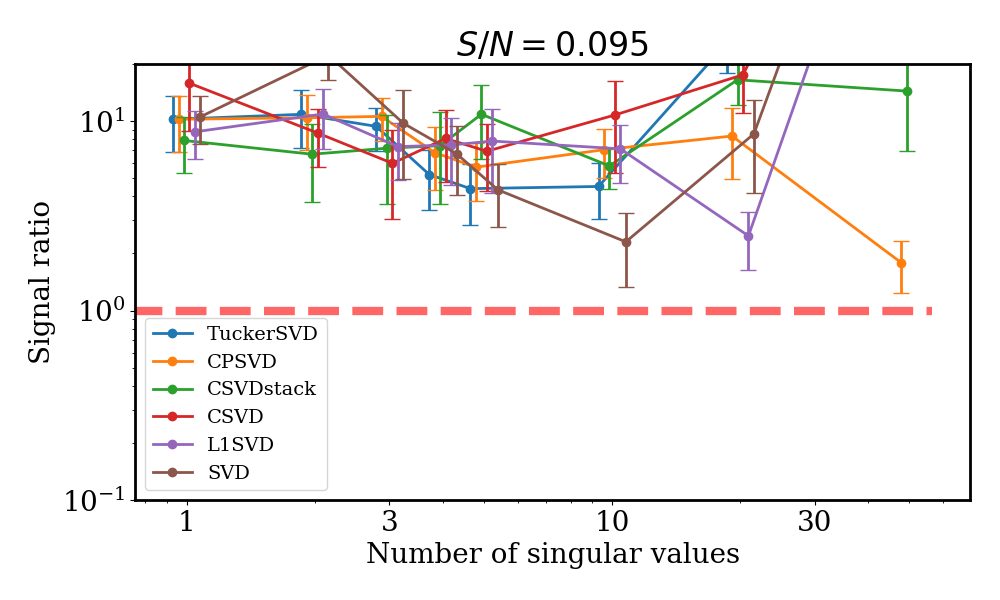}
    \caption{ {The ratio of the recovered signal to the recovered noise-free signal as a function of the number of singular values removed from the LMC/G23 dataset for the six SVD methods. Signal ratio is defined as the ratio of the cross-power spectrum to the original intensity map power spectrum, averaged over all $k$, and this is normalised by the original intensity map power spectrum with the same number of singular values removed (i.e.\ the top panel of Figure~\ref{fig:IM_signal_loss})}. A logarithmic y-axis is used here to capture the noise-biased power spectra at low S/N ratio. }
    \label{fig:IM_signal_loss_ratio}
\end{figure}

\begin{figure*}[t]
    \centering
    \includegraphics[width=0.99\textwidth]{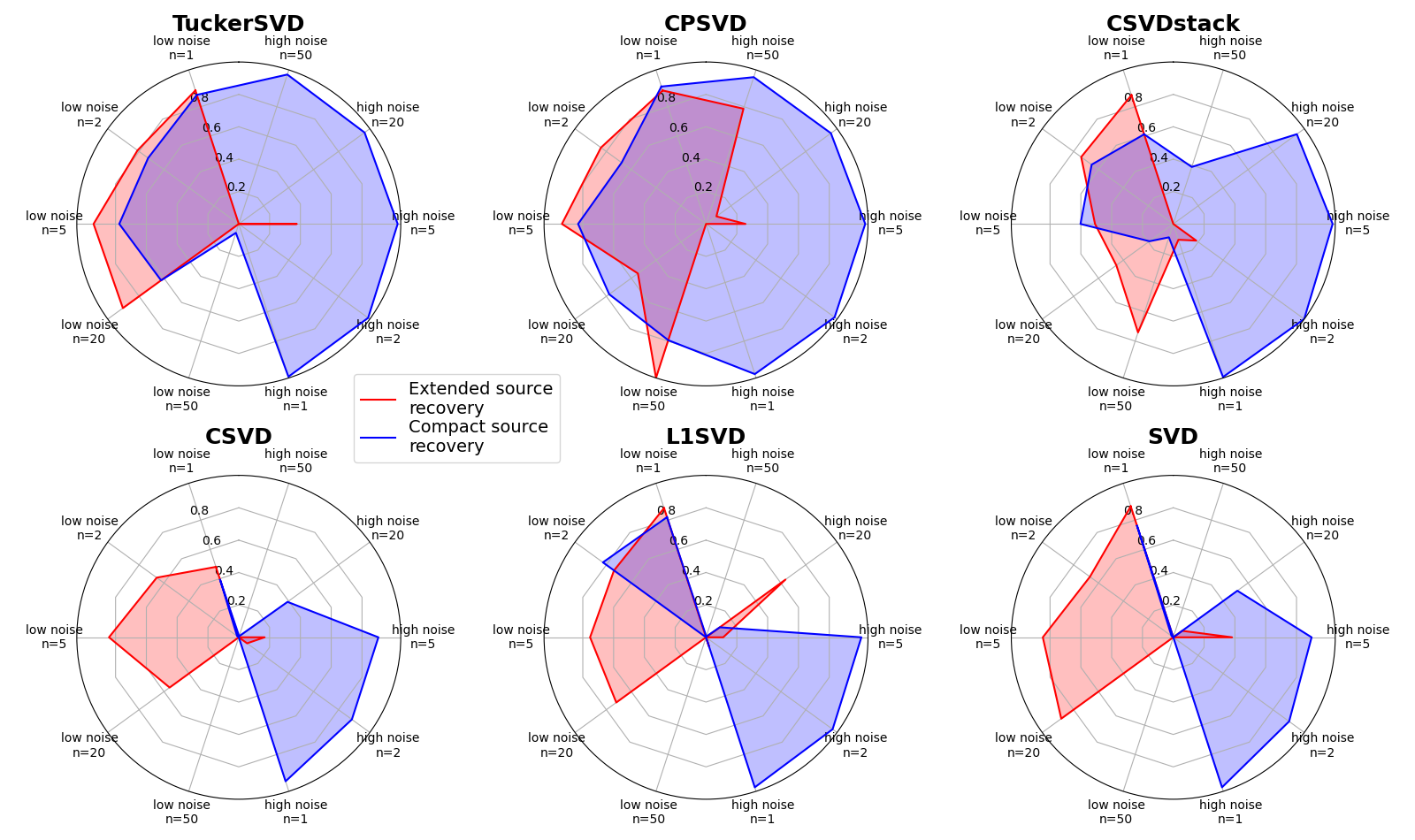}
    \caption{Radar plots showing the performance of each of the six SVD methods with respect to extracting an extended source (red) and a compact source (blue). The top row are the tensor SVD methods, and the bottom row are the 2D SVD methods. The left half of each radar plot refers to performance in a 'low noise' environment and the right hand half refers to a 'high noise' environment. For each half, we have displayed performance for different numbers of singular values: $n=$ 1, 2, 5, 20 and 50. Extended source recovery performance is judged by closeness to recovery of the mean intensity map power spectrum (logarithmic space -- see Figure~\ref{fig:IM_signal_loss_ratio}). The compact source recovery performance is from the top row of Figure~\ref{fig:signal_loss} (linear space). }
    \label{fig:IM_radar}
\end{figure*}

\begin{figure}
    \centering
    \includegraphics[width=0.99\textwidth]{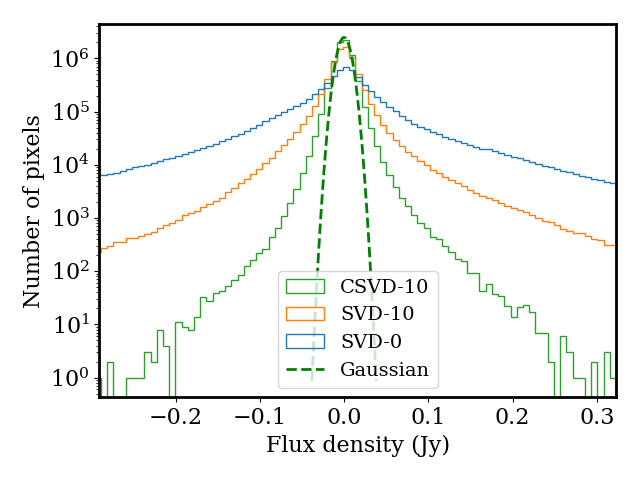}
    \caption{A histogram of the flux densities in the LMC data cube used for the compact-source injection tests. The blue histogram is the data without any singular value removal (preconditioned, but rescaled back to flux density); the orange histogram is after $n=10$ singular values removed (SVD method); and the green histogram is after clipping and $n=10$ singular values removed (CSVD method). The dashed green line is the theoretical expectation from the radiometer equation.}
    \label{fig:fluxhisto}
\end{figure}

\section{Results}
\label{sec:results}

After preconditioning and signal injection, the two example datasets were subjected to each of the six SVD methods with $1 \leq n \le 50$ singular values removed. For the three 2D techniques, $n$ is the number of singular values subtracted from each 2D frequency-time/position plane. For the 3D techniques, it is the number subtracted from the whole data cube.

\subsection{Compact sources}

An example of the progression of the flattening of the NGC 6744 dataset is shown in Figure~\ref{fig:flattening} for two example values of $n$ for the CSVD and CPSVD methods. Increased values of $n$ result in flatter backgrounds, but substantial signal loss in the case of the 2D CSVD method (the injected point source being a `wanted' signal). But, as the source is also compact in `beam space', the 3D CPSVD method is able to use the additional information to avoid signal loss and still flatten the background (with the exception of the spatially-variant component of the NGC 6744 and Galactic signals, centred at $1416.5\pm 0.8$ MHz and 1420.4 MHz, respectively).

 {For all methods, we have measured the amplitude of the recovered signal and the rms noise prior to injection for the central beam, at the position of the injected signal. 
The results are summarised in Figure~\ref{fig:signal_loss}. The recovered signal  is plotted as a function of $n$ in the top row of Figure~\ref{fig:signal_loss}, which shows that the 2D techniques (SVD, L1SVD, CSVD) lose all injected signal ($>99$\%) beyond $n=2$ in the NGC 6744 field and most signal ($>50$\%) at $n>10$ in the LMC field. In contrast, the true 3D techniques (CPSVD and TuckerSVD) only partially lose signal even beyond $n=10$ in the NGC 6744 field ($<40$\% at $n=20$), and lose even less ($<5$\%) at $n=50$ in the LMC field. }

In terms of rms residual (the middle row of Figure~\ref{fig:signal_loss}), the 2D techniques obtain a lower noise in the clean NGC 6744 field -- these techniques are fitting plane-by-plane, so are actually fitting $\sim 72/n$ times the total number of singular values compared with, say, the CPSVD method. However, in the `challenging' wideband LMC field, all methods reach a similar rms for $n>2$, although the TuckerSVD approximation starts failing at $n>20$.

The most significant plots in Figure~\ref{fig:signal_loss} are the bottom row S/N ratio comparisons - i.e.\ the ratio of the upper two rows. High values for the S/N ratio are measured in the NGC 6744 field at $n=1$ for all methods, but this drops to zero for the 2D techniques with $n>2$. In the LMC field, the S/N also falls to near-zero for the 2D techniques at $n>10$. In contrast, high S/N ratio is obtained for all values on $n$ for the 3D techniques, with CPSVD being the outstanding performer.  

\subsection{Extended sources}

The extended G23 intensity map was injected into the LMC dataset in the frequency range 1300 to 1416 MHz (corresponding to $z < 0.09$). Given the low amplitude of the G23 signal ($T_B = 52 \mu K$ at $z=0.1$; see Eq.\ref{eq:TB}), the short observing time, the strong and variable continuum in the field (up to $\sim 30$ Jy), and the RFI environment at Murriyang, recovery at this level would be too challenging a task. But artificially boosting the mock G23 map into the mK regime, so that S/N is in the range $0.1 \sim 1$, better simulates an actual observation, with all its defects, and still represents a better test for the efficacy of SVD techniques in a real observational environment, compared to using simple white-noise tests.  {Correlations between beams is retained, but genuine HI galaxy and intensity signals from behind the LMC are down-weighted, and not picked up in the cross power spectrum.} 

Examples of 2D waterfall sections of the LMC/G23 data are shown in Figure~\ref{fig:lmcflattening}, alongside the same sections following tensor SVD flattening, in this case Canonical Polyadic SVD (CPSVD) with 10 singular values removed. To demonstrate the ability of the various SVD methods to de-noise the data and potentially recover the intensity map, we employ power spectrum techniques, closely mimicking the normal manner in which the statistical properties of linear density fields are derived from cosmological measurements, such as for redshifted HI signals in the EoR \citep{2022ApJ...924...51A, 2022A&A...666A.106T} and the HI intensity mapping of large-scale structure \citep{2023MNRAS.518.6262C}. 

We define the power spectrum of the intensity map as $\Delta_i^2(k) = (2\pi)^{-2} k^3 P_i^2(k)$, where $P_i(k)$ is the spherical average of the 3D Fourier transform of the intensity map, and $k$ $(=2\pi/\lambda)$ is in units of inverse comoving Mpc according to Planck2018 cosmology \citep{2020A&A...641A...6P}. In a similar manner, the cross-power spectrum is given by $\Delta_x^2(k) = (2\pi)^{-2} k^3 P_i(k)P_c(k)$, where $P_c(k)$ is the spherical average of the 3D Fourier transform of the measured temperature field (including any mock intensity map). For the current purposes, we have not re-sampled the sky into real space, so $k$ values are only accurate at the centre point of the data cube -- in other words, corrections for sky curvature, volume etc. have been ignored. This will result in some mixing of $k$-modes, which is not particularly important here, as we will later be averaging these when estimating signal recovery.

Figure~\ref{fig:PSexamples} shows example power and cross-power spectra for the LMC/G23 dataset of Figure~\ref{fig:lmcflattening}. The spectra without any SVD correction are shown on the left, and the spectra obtained with CPSVD correction ($n=10$) are shown on the right. The error bars represent the standard error of the mean power over each bin in $\Delta_i^2(k)$ and  $\Delta_x^2(k)$. The low-$k$ offsets of the cross-power spectrum (blue points) from the true power spectrum (red points) are very large for the no-SVD cross-power spectrum. The errors are also larger than expected from sample variance. However, the CPSVD cross-power spectrum reasonably reproduces the injected G23 power spectrum, although it remains noisy at intermediate values of $k$, probably reflecting the location of the residual RFI band. The two high-$k$ values are also offset from the injected G23 power spectrum, possibly due to noise bias. 

In Figure~\ref{fig:PSnsvdgallery}, we extend this analysis to all six SVD methods and multiple singular values (the same range as for Figure~\ref{fig:signal_loss}, plus $n=0$). 
We estimate  {signal recovery} from the ratio of the cross-power spectrum to the original G23 intensity map power spectrum. 
Since the signal-to-noise ratio is low (S/N $\approx 1$), the amplitude of the measured power spectrum is generally too high (yellow/orange color) for low $n$, but quickly approaches unity (orange/red colors) at $n=3-5$ singular values for the upper tensor SVD techniques. The 2D SVD techniques (lower panels in Figure~\ref{fig:PSnsvdgallery}) reach the unity power spectrum ratio a bit more quickly, but over-subtract low-$k$ modes even more quickly, resulting in larger signal loss. All methods remove signal at the lowest value of $k$ ($\le 0.03 h$ Mpc$^{-1}$) for $n\ge 1$. As previously noted for Figure~\ref{fig:signal_loss}, the Tucker method fails for $n>20$.

A more compact representation of Figure~\ref{fig:PSnsvdgallery} is given in Figure~\ref{fig:IM_signal_loss} by averaging across all $k$-values. This enables a more quantitative comparison of extended source  {signal recovery} (defined by the average ratio of the cross-power spectra to the G23 power spectra), and better exploration of the effect of signal-to-noise ratio (defined by the ratio of the average G23 power to the cross-power spectrum in the absence of any injected signal). The upper panel shows the manner in which the average amplitude of the power spectrum of the injected intensity map reduces with the number ($n$) of singular values removed. The lower two panels show how this average changes at low S/N ratios. At S/N=0.7 (centre panel), the signal loss is fairly minor across all methods when $n=1$. But, as for compact sources, signal loss is less severe for larger $n$ for the tensor SVD methods, especially CPSVD and Tucker. At lower S/N ratio (0.095 in the bottom panel), there is large noise bias, which is only lowered at large values of $n$. However, to determine if these are real detections, we need to know how much signal as well as how much noise is being removed. If more signal than noise is being removed, then any detections at low S/N ratio are probably meaningless.

It is therefore useful to normalise the average cross-power amplitude $\bar{\Delta}^2_x (n)$  by the intensity map power spectrum $\bar{\Delta}^2_i (n)$ after applying the same SVD method and similar $n$. This result is shown in Figure~\ref{fig:IM_signal_loss_ratio}. At S/N=0.7, it is clear that the signal reduction seen in Figure~\ref{fig:IM_signal_loss} is matched by a similar noise reduction, so that signal recovery is robust for all SVD methods for $n\leq20$, but failing for the 2D methods at $n>20$. There remains a small ($\sim 30$\%) noise bias across all methods. However, at the lowest S/N=0.095 at the bottom of Figure~\ref{fig:IM_signal_loss_ratio}, only the CPSVD technique approaches zero signal loss, but only for the largest~$n$.

\section{Discussion}
\label{sec:discussion}

Our initial observations with the cryoPAF have found that it has more than met its design goals with respect to sensitivity and field of view. Placed on the Murriyang telescope at Parkes, the sub-20 K system temperature on 72 beams covering  $\sim1$~deg$^2$ at 1.4 GHz will constitute a powerful instrument for surveys of HI, OH, pulsars and FRBs. 
Careful analysis of initial HI observations show excellent agreement with previous results, and have highlighted how quickly the cryoPAF can achieve better sensitivity than the former multibeam instrument in a fraction of the observing time. A surprising consequence of our test observations with the LMC was the ease in which spectral calibration was possible -- i.e.\ no frequency-switching or high-order polynomial removal -- just a straight-line fit to zoom-band spectra. This enabled the discovery of a low-column density component ($N_{\rm HI} \approx 8\times 10^{18}$ cm$^{-2}$) of the LMC which was previously missed, presumably due to signal-loss by overfitting. 

The same datasets have also  {been} used to investigate improved methods of noise suppression for future cryoPAF observations -- particularly the removal of spectral features due to RFI and continuum sources. As demonstrated (to a small extent) for the LMC, such suppression often results in signal loss, which needs to be avoided.
To quantify noise suppression and signal loss, injection of artificial compact and/or extended sources is a powerful technique. As shown in Section~\ref{sec:results}, higher-order SVD is a powerful method for noise suppression and signal retrieval when applied to cryoPAF commissioning data. However, signal loss will always depend in detail on the nature and level of the RFI and the significance of foreground/background contamination for every observation, even at fixed number of singular values, $n$. This is because the unwanted components are usually stronger, so their removal will always come before removal of weaker RFI or residual continuum. For instruments like the cryoPAF, a wider range of commissioning data under many different circumstances (e.g.\ RFI nature, frequency, bandwidth, time of day) is still required to better understand and improve methods for noise reduction and signal retention.

Additionally, in cases such as the G23 field for $z<0.09$ (the upper redshift limit of the available commissioning data), the volume is only $3\times10^5$ cMpc$^3$, implying that there will be large cosmic variance on the power spectrum at wavenumbers of $k\leq 0.01 h$ cMpc$^{-1}$. So, whilst we can make accurate estimates of signal loss for injected signals, we cannot do this for the same accuracy for cosmological signals in volumes small enough to be affected by cosmic variance.  

Nevertheless, our results establish some ground truths around signal loss and noise reduction for cryoPAF data, and possibly other datasets, under a range of observing conditions and source characteristics. We therefore attempt to consolidate the results of Section~\ref{sec:results} into a representative set of figures of merit with which to compare tensor SVD, SVD and other de-noising methods in the future. We have combined the results from Section~\ref{sec:results} into a set of six radar plots shown in Figure~\ref{fig:IM_radar}. These plots combine the results from sources of different spatial and spectral extent, different noise levels, and different depths of singular value removal. The usual radar plot normalisation of good = 1 and bad = 0 is used, with the metrics being directly related to signal loss in Figure~\ref{fig:signal_loss} (top row) for compact sources, and normalised signal loss in Figure~\ref{fig:IM_signal_loss_ratio} for the extended sources.

For compact sources (blue polygon), the largest radar coverage is offered by the tensor SVD techniques in the top row of Figure~\ref{fig:IM_radar}, with the true tensor techniques (Tucker and CPSVD) being the `roundest' -- i.e. the highest efficacy across different noise and SVD depth regimes. Helped by compactness in the frequency domain, the 2D SVD techniques (bottom row) do almost as well at signal recovery in the high-noise regimes. Paradoxically, apart from $n=1$, they fail in the low-noise regimes because of signal loss -- the most significant singular values are related to the source itself. In our example dataset, signal at $n=1$ is only saved because of the presence of the NGC 6744 signal, which is the first to be subtracted.

For extended sources (red polygons), the largest radar coverage is offered by the TuckerSVD and CPSVD techniques, but the L1SVD technique works surprisingly well for $n=20$ in the high-noise regime. This is possibly because L1 may be more appropriate than clipping due to the substantial non-Gaussianity in our example dataset. However, this comes at the significant expense of processing speed. 
Nevertheless, of all the example methods, the best extended source performance in the low and high-noise regimes is offered by the high-$n$ CPSVD method.

Overall, the tensor SVD methods show a lot of promise for de-noising data of  cryoPAF data. This is particularly highlighted for sources which are compact in  {all} dimensions, but in general seems to be true for sources, such as the G23 intensity map, which have power at all scales in all dimensions. 

Regarding outliers, the tests presented here have mostly been made using robust (L1SVD), clipped (CSVD, CSVDstack) or censored (TuckerSVD, CPSVD) algorithms. Only the basic SVD method uses fully least-squares L2 methodology. The reason for the emphasis on robustness is that post-EoR cosmological HI signals are typically at the $\sim 0.1$ mK level, whereas diffuse Galactic foreground, away from the Galactic Plane itself, is at the few K level, and `regular' RFI can be at the MJy level \citep{2022PASA...39...26S}. Such high dynamic ranges are made possible with modern radio telescopes due to receiver linearity and the use of hierarchical polyphase filterbank spectrometry. Unsurprisingly, as seen in the right hand column of Figure~\ref{fig:signal_loss} and the lower row of the radar plots in Figure~\ref{fig:IM_radar}, the L1SVD method seems to be the best of the 2D techniques in terms of compact and extended source recovery. However, the differences between SVD and its clipped counterpart, CSVD, are more minor, implying that removal of outliers does not much affect the SVD solution once the data are robustly scaled as part of the pre-conditioning process. Probably this is due to the outliers in these datasets being mostly foreground or RFI vectors which are linearly correlated and therefore automatically dealt with by SVD. But it may also be due to the limited nature of the clipping (0.5 and 99.5 percentiles) and the substantial residual non-Gaussianity -- see Figure~\ref{fig:fluxhisto}. Nevertheless, the success of the L1SVD approach relative to other 2D techniques does imply that investigation of robust L1 tensor SVD techniques may be useful for higher large multidimensional datasets \citep[e.g.][]{doi:10.1137/23M1574282, 10598112}, and therefore a possible alternative to the successful CPSVD method to explore.

\section{Conclusions}
\label{sec:conclusions}
We have presented the first spectral-line commissioning data from the  Murriyang cryogenic phased array feed (cryoPAF). Its performance is verified through calibrator observations, and observations of nearby galaxies. Our findings have been:

\begin{itemize}

\item The system temperature at 1.4 GHz has been found to be excellent. For beams with 0.3 deg of the optical axis, values of its ratio of with dish aperture efficiency and main beam efficiency were found to be $T_{\rm sys}/\eta_{\rm d} = 25.5$ K and $T_{\rm sys}/\eta_{\rm mb} = 21.1$ K, respectively. Combined with earlier hot-cold load measurements implying $T_{\rm sys} = 17$ K \citep{10238280}, efficiency values of $\eta_{\rm d} \approx 0.7$ and $\eta_{\rm mb} \approx 0.8$ are derived.

\item Zoom band spectra of NGC 6744 and the LMC show good agreement with previous observations in velocity and flux density. Aside from times when the digital filterbank flattening module was switched off, spectra were extremely flat. A complete map was made of the LMC with a $7\times 7$ pointing grid in `camera' mode. This enabled a thorough investigation of performance relative to previous observations. Firstly, the measured noise ($\sim 18$mK) was consistent with the observational parameters (i.e.\ system temperature, integration time, frequency resolution and beam overlap). This implies an rms column density sensitivity of $3.6\times10^{16}$ cm$^{-2}$ in the 1.1 km~s$^{-1}$ channels of the gridded cube. Secondly, comparison with previous multibeam observations (which have a higher rms despite $30\times$ more on-source integration time) shows excellent temperature agreement. The only scale difference was attributable to different calibrator brightness temperature values. However, a low level of emission  {($0.1\sim 0.3$~K)} was picked up in the cryoPAF data, but not seen in the multibeam data. This is attributed to signal loss due to high-order polynomial subtraction of spectral and spatial baselines required to process the narrow-band frequency-switched multibeam data. On the other hand, only  {linear baseline fits} in frequency space were required for the cryoPAF data.

\end{itemize}

A major part of this study has also been to use the commissioning data to test possible future techniques for the retrieval of weaker signals from compact and extended sources, focusing on the suppression of continuum source contamination and radio-frequency interference (RFI). We measured the effect on signal loss and signal-to-noise ratio of various suppression techniques, focusing on the well-proven technique in linear algebra of singular value decomposition (SVD). This method has been widely used for foreground suppression in previous cosmological studies of redshifted 21-cm atomic hydrogen but, due to the nature of the observational data, mainly in its two-dimensional matrix or unfolded tensor mode. In this paper, we have extended these studies to higher order, commensurate with the high dimensionality of cryoPAF data. The main findings of this aspect of our study are:

\begin{itemize}

\item  {Robust pre-conditioning of the data was found to be essential for stable results and algorithm convergence. Further robustness (clipping or L1 norm) was useful in greatly reducing rms residuals, but not necessarily advantageous in aiding signal recovery.}

\item For compact sources, 2D SVD techniques are effective in suppressing noise, but are also effective in suppressing signal unless care is taken to understand signal loss as a function of RFI and foreground environment. Source flux density can be completely removed for $n = 1$ or $2$ in our `clean'  {NGC 6744} ndataset, and for $n \geq 20$ for our `challenging' dataset  ($n$ is the number of singular values removed). 

\item For compact sources, higher-order or tensor SVD techniques are more  {robust} than 2D SVD, providing lower signal loss and higher signal-to-noise ratio over a larger range of singular values. These methods have the obvious advantage of being able to utilise source compactness in the third dimension. This benefit is shared, but less effectively, with the `stacked' SVD technique (CSVDstack, in this study), where the third dimension of time/position is unfolded onto the second dimension. 

\item For extended sources, we examine signal loss using power spectrum techniques, in common with methods normally employed in the search for cosmological signals. We use an `intensity map' generated from the GAMA G23 galaxy survey. At `high' signal-to-noise ratios (S/N $\sim 1$), the 2D techniques (especially the robust L1SVD technique) allow convincing signal recovery, although suppress power at small wavenumbers, or large scales ($2\pi/hk \geq 300$ cMpc) for the small volumes tested here. However, at low signal-to-noise ratios (S/N $\sim 0.2$), only the tensor SVD techniques permit source recovery.

\item Of the tensor SVD techniques, the clipped Canonical Polyadic decomposition (CPSVD) method has better signal recovery and signal-to-noise ratio than the Tucker decomposition for noisy data.

\end{itemize}

Finally, there is a strong relationship between tensor SVD and deep learning techniques, with the latter now having a growing literature in the context of cosmological studies. Deep learning is a promising method for better identifying outliers, dealing with non-linear dependencies, and recovering signal in spaces of high dimensionality. However, substantial groundwork will be required in training, tuning and transfer techniques to best utilise its advantages in the face of complex RFI environments, limitations in antenna calibration and complex foreground radiation fields.

\begin{acknowledgement}

Murriyang, CSIRO’s Parkes radio telescope, is part of the Australia Telescope National Facility (https://ror.org/05qajvd42) which is funded by the Australian Government for operation as a National Facility managed by CSIRO. We acknowledge the Wiradjuri people as the Traditional Owners of the Observatory site. The cryoPAF was funded by the Australian Government through CSIRO and the Australia Research Council (project LE200100096).

This work made use of Python, numpy \citep{harris2020array}, SciPy \citep{2020SciPy-NMeth}, TensorLy \citep{JMLR:v20:18-277}, h5py, matplotlib \citep{2007CSE.....9...90H} and Astropy,\footnote{http://www.astropy.org} a community-developed core Python package and an ecosystem of tools and resources for astronomy \citep{astropy:2013, astropy:2018, astropy:2022}.

\end{acknowledgement}

\bibliography{main.bib}

\end{document}